\documentclass[twoside,english]{pnreport}
\usepackage[T1]{fontenc}
\usepackage[latin1]{inputenc}
\usepackage{verbatim}
\usepackage{floatflt}
\usepackage{graphicx}
\usepackage{amssymb}
\usepackage[dvipdfm]{hyperref}

\usepackage[hyperref,rgb]{xcolor}

\makeatletter

\usepackage{wrapfig}
\usepackage{rotating}
\usepackage{array}

\usepackage{babel}
\makeatother
\begin{document}

\newcommand{\shadedfigure}[2][tbp]
{{ \begin{figure}[#1]
   \colorbox[rgb]{1,1,0.92}{
     \begin{minipage}[t]{1\columnwidth}
      #2
     \end{minipage} 
    }
    \end{figure}
 }
}

\title{CT10 parton distributions\\ and other developments in the global QCD
analysis}

 \author{Marco Guzzi$^1$, Pavel Nadolsky$^1$, \\\vspace{0.1cm}
Edmond Berger$^2$, 
Hung-Liang Lai$^3$, Fredrick~Olness$^1$, and C.-P.~Yuan$^4$\\\vspace{0.1cm}
$^1$ Department of Physics, Southern Methodist University, Dallas, TX 75275 USA\\\vspace{0.1cm}
$^2$ High Energy Physics Division, Argonne National Laboratory, Argonne, Illinois 60439, USA\\\vspace{0.1cm}
$^3$ Taipei Municipal University of Education, Taipei, Taiwan\\\vspace{0.1cm}
$^4$ Dept. of Physics \& Astronomy, Michigan State University, E. Lansing, MI 48824, USA}

\maketitle

\begin{abstract}
We summarize several projects carried out 
by the CTEQ global analysis of parton distribution
functions (PDFs) of the proton during 2010. We discuss a recently released
CT10 family of PDFs with a fixed and variable QCD coupling strength; implementation of
combined HERA and Tevatron lepton asymmetry data sets; theoretical issues 
associated with the analysis of $W$ charge asymmetry in PDF fits; 
PDFs for leading-order shower programs;
and constraints on new color-octet fermions from the hadronic data. 

\end{abstract}

\begin{center}
\date{\today}
\end{center}

\tableofcontents
\newpage

\subsection{CTEQ PDFs in 2010}
Parton distribution functions (PDFs) are essential nonperturbative
functions of quantum chromodynamics (QCD). They describe 
the internal structure of the proton in high-energy scattering
at the Fermilab Tevatron collider, CERN Large Hadron
Collider, and in other experiments. Modern PDFs continuously evolve 
to include emerging 
theoretical developments and latest data from hadronic
experiments, and to provide reliable estimates of uncertainties
associated with  various experimental and theoretical inputs. In this
paper, we review the recent progress in the determination of the PDFs by  
CTEQ collaboration \cite{Pumplin:2009nk,Lai:2010nw,Lai:2010vv,Lai:2009ne}, which
is one of three groups involved in the global analysis of hadronic
data, besides the MSTW \cite{MSTW08} and NNPDF \cite{NNPDF21}
groups. 

\subsection{Implementation of new data sets}
Since the release of the previous general-purpose CTEQ6.6 PDF set
\cite{Nadolsky:2008zw} in 2008, 
new data sets have been published in every category of processes
included in the global QCD analysis: deep inelastic
scattering (DIS), vector boson production, and inclusive jet
production. These data sets include a combination of DIS cross
sections by the H1 and ZEUS collaborations in HERA-1 \cite{2009wt}, as
well as measurements of $W$ lepton asymmetry \cite{Acosta:2005ud,d0_e_asy,d0_mu_asy}, $Z$ rapidity
distributions  \cite{Aaltonen:2010zza,Abazov:2007jy}, and single-inclusive jet cross sections
\cite{Aaltonen:2008eq,:2008hua} by CDF and D\O\  
collaborations at the Tevatron. All these new data 
are included in our latest global analysis, designated 
as CT10 \cite{Lai:2010vv}. 

The new analysis produced two families of 
general-purpose PDF sets, denoted as CT10 and CT10W,  which differ in their
treatment of the Tevatron $W$ lepton asymmetry data sets affecting the
ratio of $d$ and $u$ quark PDFs at $x > 0.1$, as discussed below, but
are very similar in all other aspects. In addition, we examined the
dependence of the PDFs on the QCD coupling $\alpha_s(M_Z)$ and 
provided special CT10AS PDF sets with a varied $\alpha_s(M_Z)$ in the range
0.113-0.123 to evaluate the combined PDF-$\alpha_s$ uncertainty in
practical applications.  The CT10 PDFs are obtained at
next-to-leading order in $\alpha_s$, using the general-mass treatment
of charm and bottom quark contributions to hadronic observables.
To support calculations for heavy-quark production in 
the fixed-flavor-number factorization scheme, 
we also provide additional PDF sets CT10(W).3F and CT10(W).4F, 
obtained from the best-fit CT10.00 and CT10W PDF sets by QCD evolution 
with three and four active quark flavors. 
All the PDF sets discussed in this paper (CT10, CT10W, CT10AS, CT10XF,
and CT09MC) are available as a part of 
the LHAPDF library \cite{lhapdf} and from our website ~\cite{website}.

\subsection{Constraints from combined DIS data by HERA-1} 
The CT10/CT10W fits include a combined set of HERA-1 cross sections 
on neutral-current and charged-current DIS \cite{2009wt}, 
which replaces 11 separate HERA-1 data sets used in
CTEQ6.6 and earlier fits. In the combined set, 
systematic factors that are in common to both experiments 
were presented as a table of 114 correlated systematic errors,
whose effect is shared by each data point in all scattering channels.
As a result of the cross calibration of detection parameters 
between the H1 and ZEUS experiments, the combined data set has a reduced
total systematic uncertainty. Consequently, the 
PDF uncertainties at $x < 10^{-3}$, in the region where the HERA data
provide tightest constraints on the gluon and heavy-quark PDFs, 
are also reduced. 

The impact of the combined HERA-1 set on the PDFs is illustrated by
Fig.~\ref{CT10changes}, showing relative differences between the CT10 PDF
set, fitted to the combined HERA-1 data, and a counterpart fit, fitted
to the separate HERA-1 data sets. In the left subfigure, comparing the
best-fit PDFs in the two fits, one observes reduction 
in the gluon and charm PDFs at $x < 0.05$, accompanied by a
few-percent increase in the $u$ and $d$ quark PDFs in the same $x$
region. The strange quark PDF shows a larger suppression (up to $25\%$
at $x=10^{-5}$), which, however, is small compared to the large PDF
uncertainty associated with this flavor. 
Fig.~\ref{CT10changes} (right) shows the asymmetric
fractional PDF uncertainty, computed as in \cite{Nadolsky:2001yg},
and normalized to the best-fit gluon PDF of each fit.
The impact of the HERA-1 data on the uncertainties
of the gluon and charm PDFs is visible in the small-$x$ region,
starting from $x=10^{-3}$ and going down to $x=10^{-5}$, where 
the error bands contract upon the combination of the HERA data sets. 
In the large $x$ region, the error
bands for the combined and separate HERA data sets are
almost coincident.

\shadedfigure{
\centering
\caption{\small{Left: ratios of CT10 central PDFs fitted to the
    combined HERA-1 data set and to the separate HERA-1 data sets, at scale $\mu=2$ GeV.
Right:  bands of the PDF uncertainty (relative to the the central PDF
    set) for the gluon PDF in  (red) CT10 with the combined HERA-1 data, 
(blue) CT10 with the separate HERA data. $g(x)$, $Q=2$ GeV.}} 
\label{CT10changes}
\includegraphics[width=5cm, angle=-90]{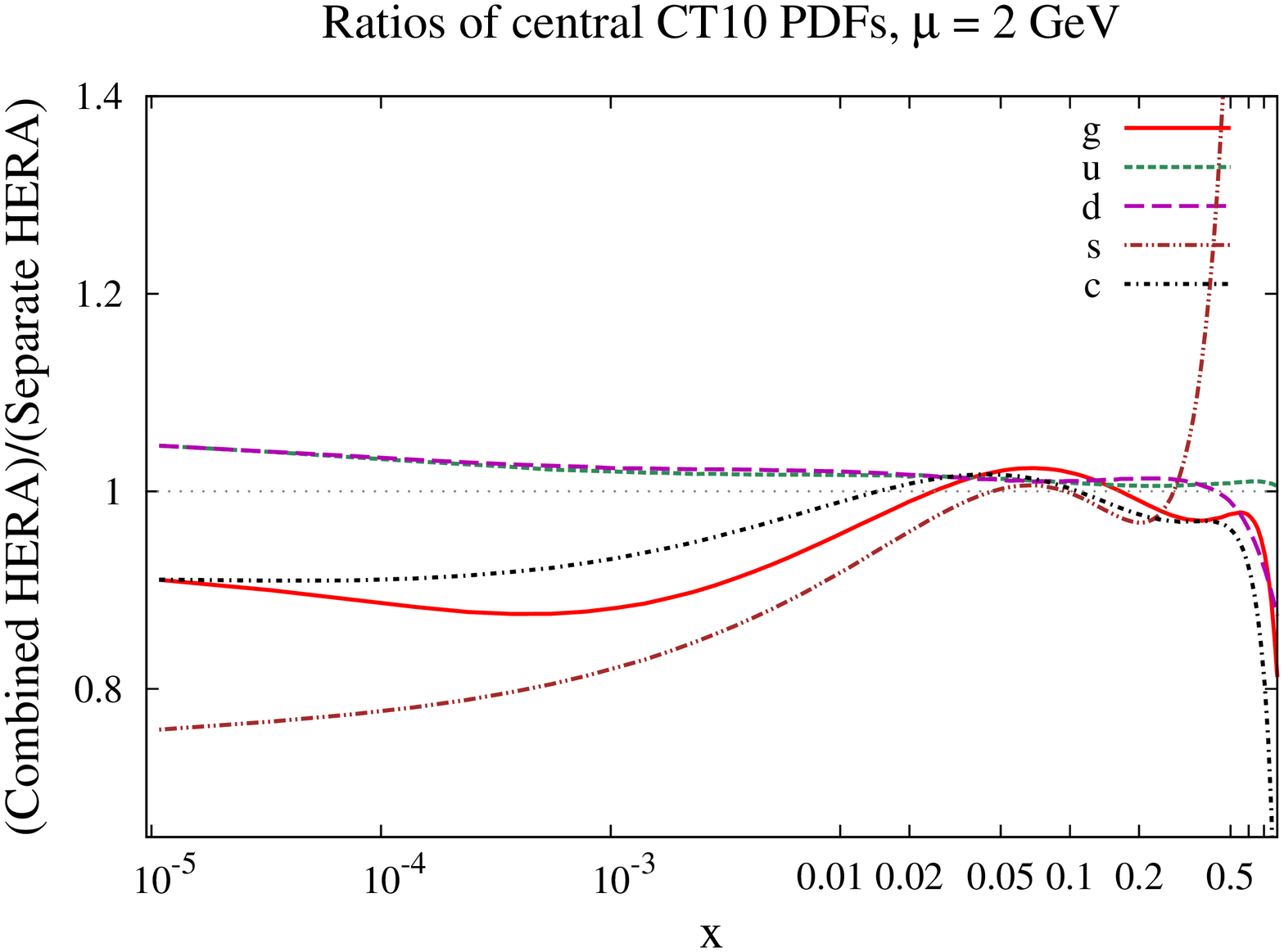}
\includegraphics[width=5cm, angle=-90]{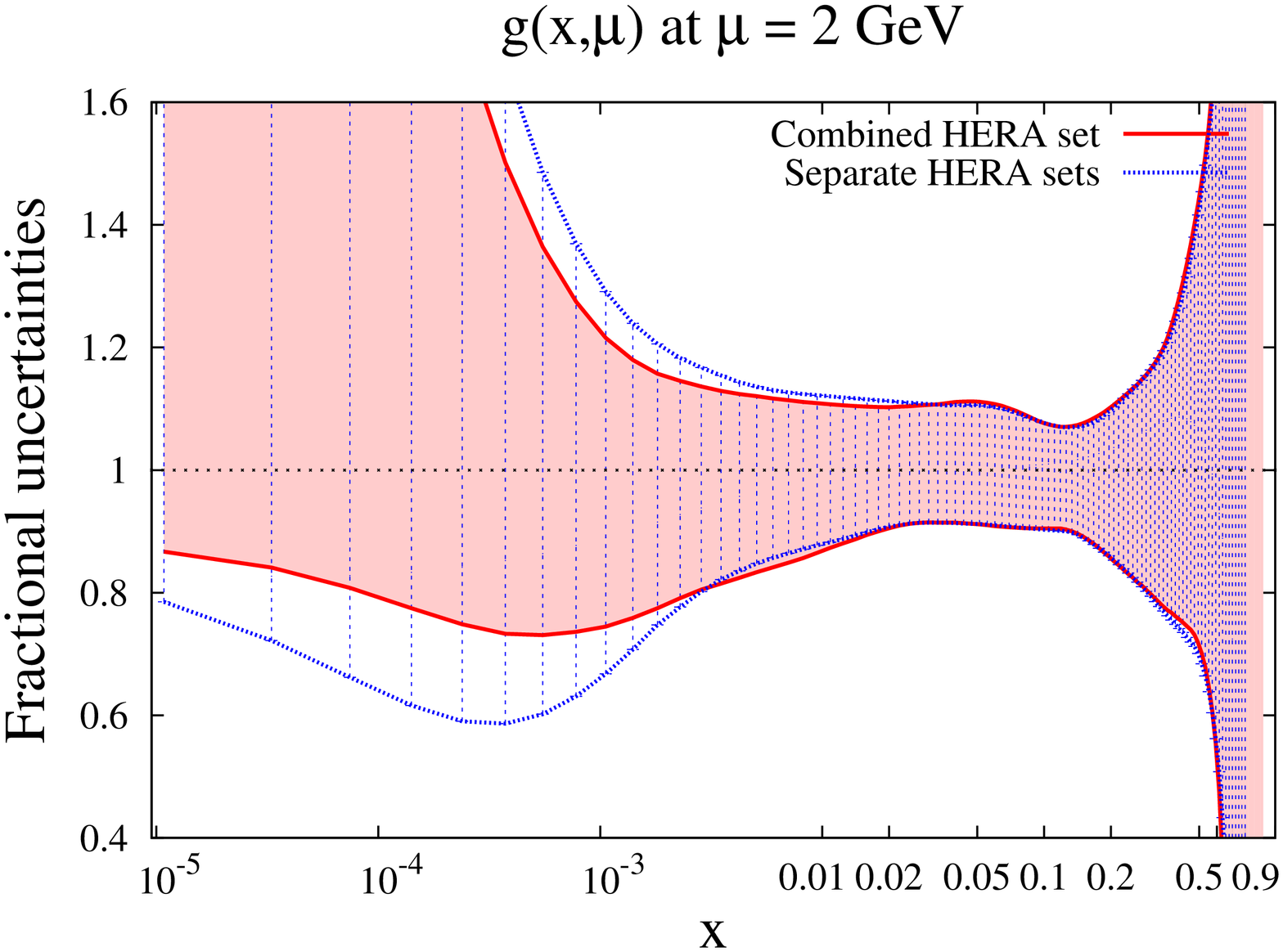}
}

\subsection{Agreement with the HERA data at small and large $x$} 
The overall 
agreement of the CT10 fit with the combined HERA-1 data is slightly worse than
with the separate HERA-1 data sets, as a consequence of some increase
in $\chi^2/\mbox{d.o.f.}$ for the neutral-current DIS data at $x<0.001$ and
$x>0.1$. While the origin of this increase is uncertain, the pattern
of point-to-point contributions to $\chi^2$ from the data 
is consistent with random fluctuations that turn out to be larger 
than normally expected. No systematic 
discrepancies between the HERA-1 DIS data and theoretical cross
sections are observed, suggesting that
the NLO QCD theory based on CT10 PDFs is generally consistent with the
HERA experiments in the region $Q>2 \mbox{ GeV}$ included in the CT10 fit. 

In looking for potential systematic deviations of this kind, we
examined the agreement with the data as a function
of either $x$ and $Q$, or the ``geometric
scaling'' variable $A_{gs}=Q^2 x^{0.3}$ proposed by NNPDF authors 
in Refs.~\cite{Caola:2009iy,Caola:2010cy}. At $A_{gs}\rightarrow 0$,
DGLAP factorization that is required to introduce the PDFs can be 
invalidated by higher-twist terms or saturation; the
question is whether such effects may bias the determination of the
PDFs at $A_{gs}\gtrsim 0.1$, in the kinematical region 
commonly included in the global fits. 

Indeed, the NNPDF study finds that the PDFs fitted to the HERA data
above some cutoff value ($A_{gs} > A_{cut}$), disagree at the
$2\sigma$ level with the HERA data in the ``causally connected''
region below the cutoff, $0.5< A_{gs} < A_{cut}$. 
The NNPDF analysis is realized 
in the zero-mass approximation and includes DIS data in the 
less safe region $\sqrt{2}\mbox{ GeV} < Q < 2$ GeV.
We repeated the $A_{cut}$ fits proposed by NNPDF as closely as
possible, in the general-mass factorization scheme, and in the region 
of $Q > 2\mbox{ GeV}$ where our data are customarily selected to
suppress higher-order and higher-twist terms. 
While the outcomes of our $A_{cut}$ fits bear some similarity to 
those by NNPDF, the discrepancies between our best-fit NLO predictions
and the data below $A_{cut}$ are less
significant than those quoted by NNPDF and are characterized by a large
PDF uncertainty. Thus, our fits do not corroborate the existence 
of stable deviations of the NLO DGLAP factorization from the data, if
the lower $Q$ bound is chosen to be above 2 GeV. See
further discussion in the appendix of Ref.~\cite{Lai:2010vv}.

\subsection{Tevatron Run-2 $W$ lepton asymmetry data}
{\bf The puzzle of Run-2 $W$ asymmetry.} 
Recently, the Fermilab D\O~ Collaboration \cite{d0_e_asy, d0_mu_asy}
published measurements of $W$ charge asymmetry $A_\ell(y_\ell)$ 
in electron ($\ell=e$) and muon ($\ell=\mu$) decay channels, presented
as a function of the rapidity of the charged decay lepton.  
NLO predictions based on CTEQ6.1 and CTEQ6.6 sets disagree with these data 
at surprisingly large $\chi^2/Npt$ of about 5. The values of $\chi^2/Npt$ can be even higher 
(as high as 20) for some 
other recent (N)NLO PDF sets \cite{d0_mu_asy,Catani:2010en}.
Such level of disagreement may appear surprising, given that 
the Tevatron $W$ asymmetry probes the ratio of $d$ and $u$ quark PDFs 
\cite{Berger:1988tu} in the region $x>0.1$, where they are known quite well from the other experiments.

{\bf Sensitivity to the $d/u$ slope.} 
The discrepancy involving $A_\ell$ can be understood in part by noticing
that the $A_\ell$ measurement is very sensitive to the average $x$ derivative
(slope) of the ratio of the up and down quark PDFs, $d(x, M_W)/u(x,M_W)$, 
computed between the typical $x$ values $x_{1,2}= M_W e^{\pm y_W}/\sqrt{s}$ accessible 
at a given boson rapidity $y_W$ \cite{Berger:1988tu,Martin:1988aj}. 
Small variations of the $d/u$ slopes in 
distinct PDF sets can change the behavior of $A_\ell$ by large amounts~\cite{Lai:1994bb}. 

\shadedfigure[tb]{
\centering
\caption{Comparison of CDF Run-2 lepton asymmetry data
  \cite{Acosta:2005ud} with LO, NLO, and resummed predictions from ResBos \cite{Nadolsky:2008unp}.}
\label{fig:WasyResum}
\includegraphics[width=0.47\textwidth]{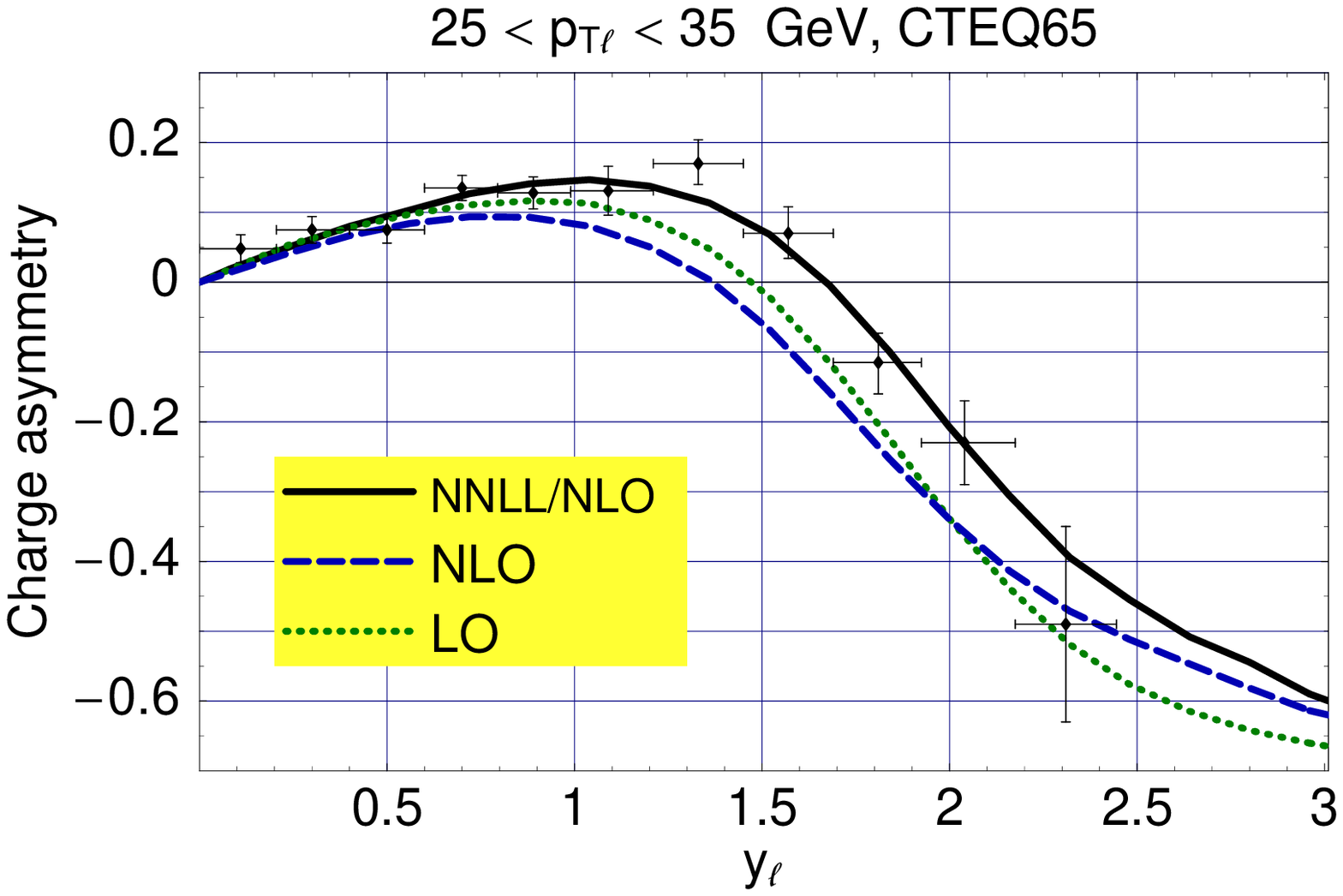}\includegraphics[width=0.47\textwidth]{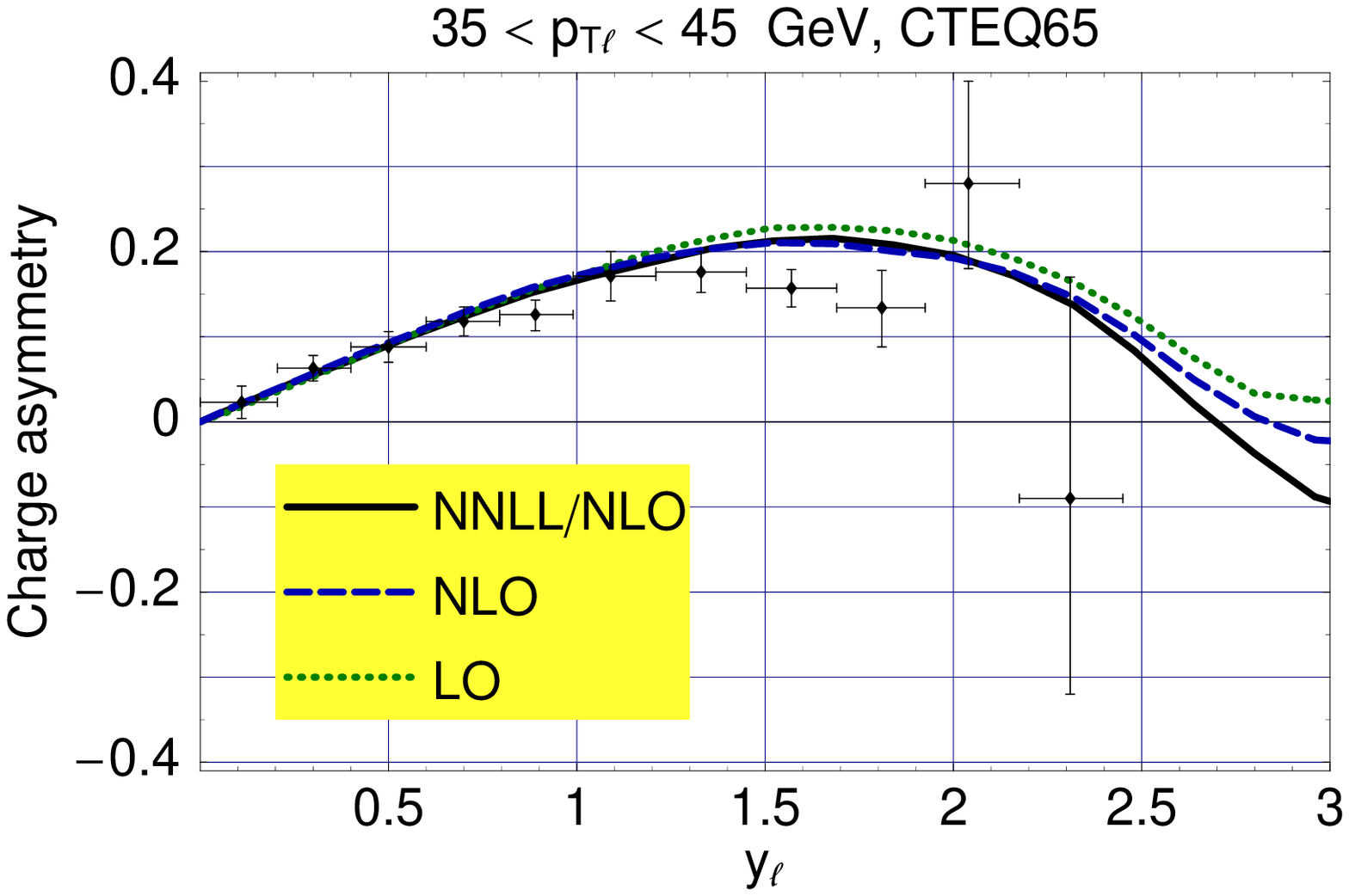}
}

\shadedfigure[p]{
\caption{Comparison of the CT10W and CTEQ6.6 predictions with the 
D\O~ Run-II data for the electron charge asymmetry $A_e(y_e)$  for 
an integrated luminosity of 0.75 ${\rm fb}^{-1}$ \cite{d0_e_asy}}
\label{figs:eleAct10w}
\centering
\includegraphics[clip,scale=0.7]{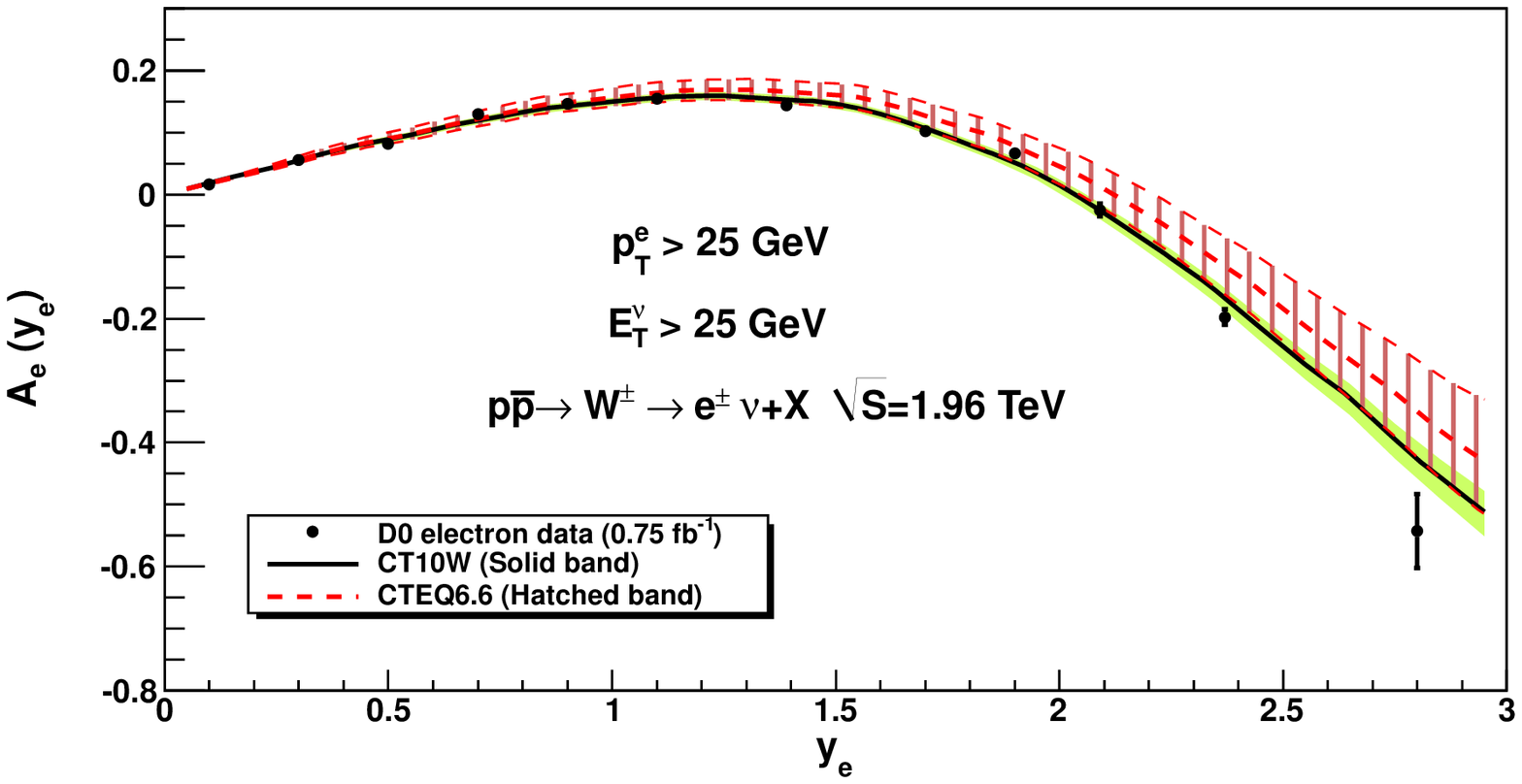} 
\includegraphics[clip,scale=0.7]{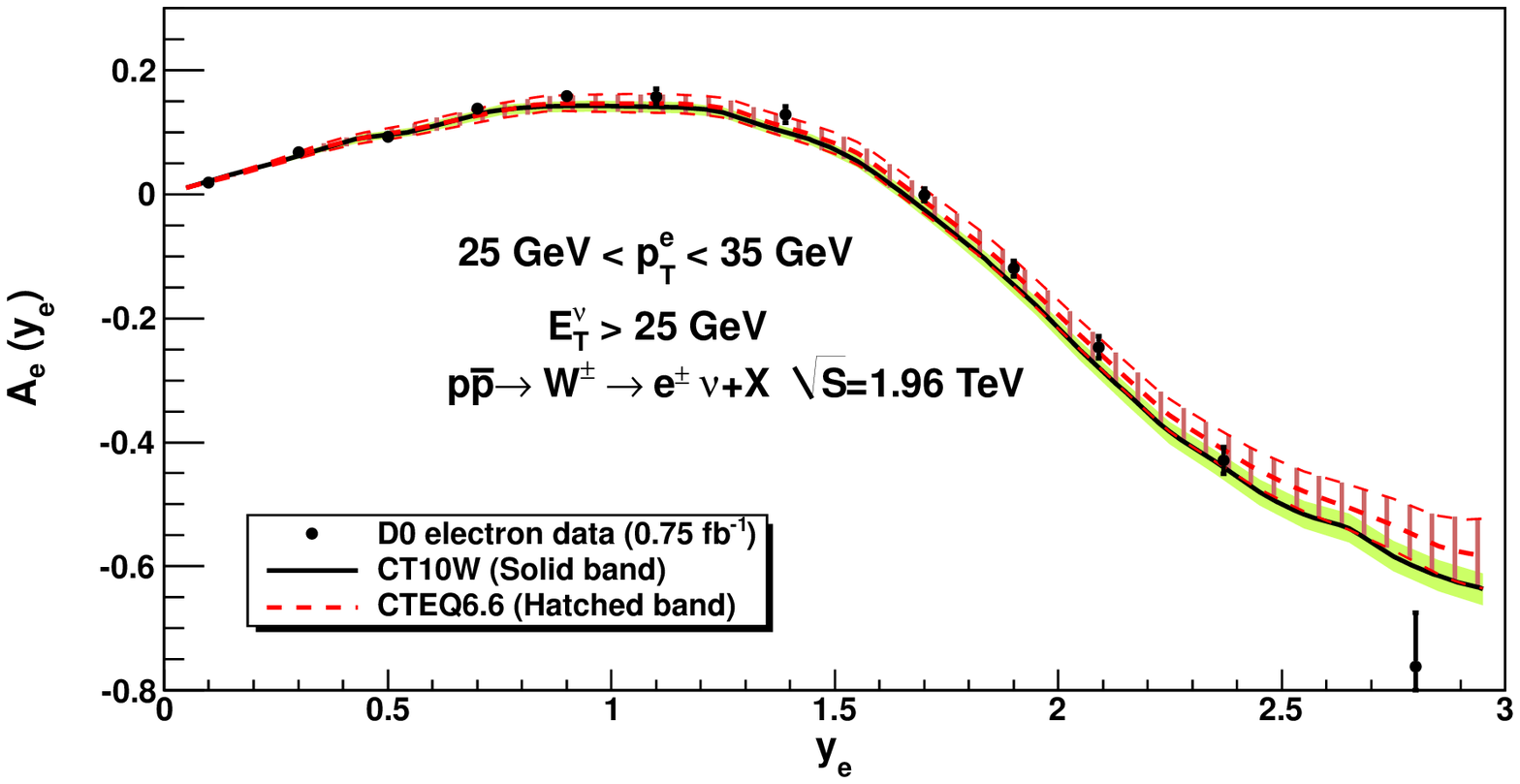}
\includegraphics[clip,scale=0.7]{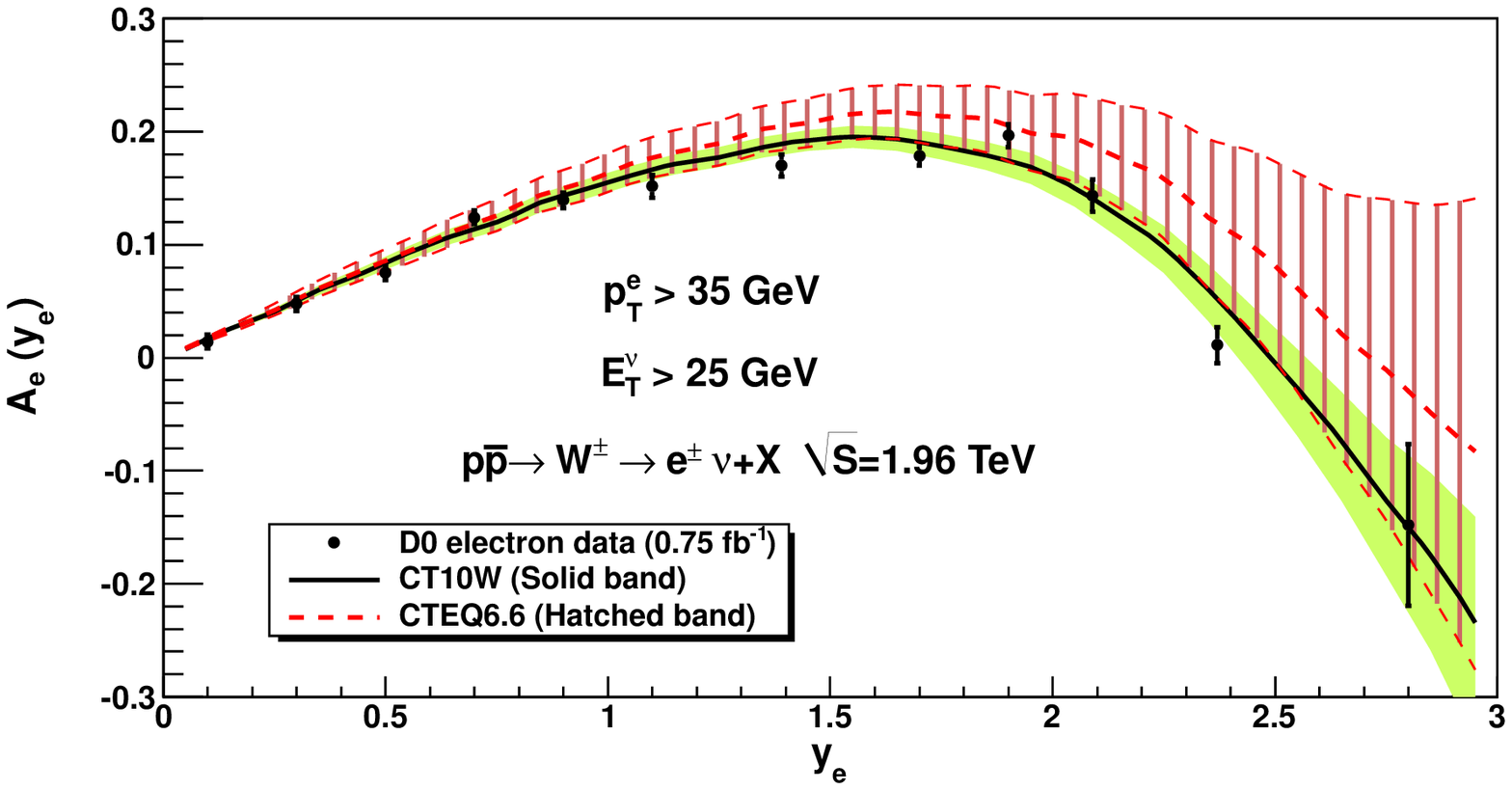} 
}

{\bf Impact of soft gluon resummation.} Another factor at play are
soft parton emissions with small transverse momenta, which affect the
precise $A_{\ell}$ data because of constraints imposed on 
the transverse momentum $p_{T\ell}$ of the decay charged lepton.
The $A_\ell$
data require $p_{T\ell}$ to be above 20-25 GeV in order to suppress
charged leptons from background processes that do not involve a $W$
boson decay. In addition, the Run-2 $A_\ell$ data are organized into
bins of $p_{T\ell}$, {\it e.g.}, 25-35 GeV and 35-45 GeV in order 
to better probe the $x$ dependence of $d(x,Q)/u(x,Q)$
\cite{Acosta:2005ud}. 
While such binning amplifies the sensitivity of the $A_\ell$ data 
to the PDFs, it also makes it dependent on the shape of the $p_{T\ell}$ 
distribution near the Jacobian peak at $p_{T\ell}\approx M_W/2 \approx
40$ GeV or, equivalently, to
the transverse momentum ($Q_T$) distribution of $W$ bosons at $Q_T < 20$
GeV, where large logarithms $\ln(Q_T/Q)$ dominate the cross
section. A calculation that evaluates these logarithms to
all orders in $\alpha_s$ \cite{Balazs:1995nz,Balazs:1997xd,Landry:2002ix}, in
addition to including 
the leading NNLO corrections \cite{Arnold:1988dp,Arnold:1989ub}, results in somewhat
different predictions for $A_{\ell}$ than (N)NLO calculations 
without resummation, like those implemented in the other available codes
\cite{Giele:1993dj,Melnikov:2006kv,Gavin:2010az,Catani:2009sm}.

In CT10 fits, the QCD radiative contributions to
$A_\ell(y_\ell)$ are implemented to the next-to-next-to-leading
accuracy in $Q_T$ logarithms and NLO accuracy in the QCD coupling strength
using the program ResBos that realizes the approach 
of Refs.~\cite{Balazs:1995nz,Balazs:1997xd,Landry:2002ix}. 
The resummed differential distributions for $d\sigma/dQ_T$ 
and $d\sigma/dp_{T\ell}$ both agree
well with the data, in contrast to the fixed-order results.  
We thus expect that the resummed predictions for $A_\ell$ implemented
in the CT10 fit are more reliable as well.

The magnitude of differences between the NLO and resummed predictions
is illustrated by Fig.~\ref{fig:WasyResum}, comparing the CDF Run-2 $A_\ell
(y_\ell)$ data \cite{Acosta:2005ud} with LO, NLO, and
resummed NNLL-NLO predictions from ResBos. 
The NLO and resummed curves are clearly distinct in the bin $25 <
p_{T\ell} < 35$ GeV, shown in the left panel, and some differences are also seen
in the bin $35 < p_{T\ell} < 45$ GeV, shown in the right panel. 
The shape of the NLO prediction in this comparison is not unique and 
depends on the phase space slicing parameter $Q_T^{sep}$ that defines 
the size of the lowest $Q_T$ bin where the real and virtual
NLO singularities are canceled \cite{Balazs:1997xd}. 
In the current comparison, 
$Q_T^{sep}=3 GeV$, but other values of $Q_T^{sep}$ are equally
possible and would lead to NLO predictions lying closer to, or further
from, the shown resummed curve. Such variations due to $Q_T^{sep}$ or
factorization scale indicate that resummation effects are important
and should be included in precise fits to $A_\ell$.\footnote{The NNLO contributions produce marginal
  modifications compared to the NNLL-NLO result included in the CT10
  analysis. We examined these contributions by redoing the calculation 
  for $A_\ell(y_\ell)$ after adding the exact $\alpha^2_s$ correction 
  for $W$ bosons produced with non-zero transverse momentum, 
  which captures a large part of the full NNLO effect. The 
changes were found to be small and inconsequential in the current fit.}

\shadedfigure[tb]{
\centering
\caption{The  $d/u$ ratio for CT10 (left) and CT10W (right) versus that for CTEQ6.6, 
at scale $\mu=85$ GeV.}
\label{figs:ct10doveru}
\includegraphics[width=0.47\textwidth]{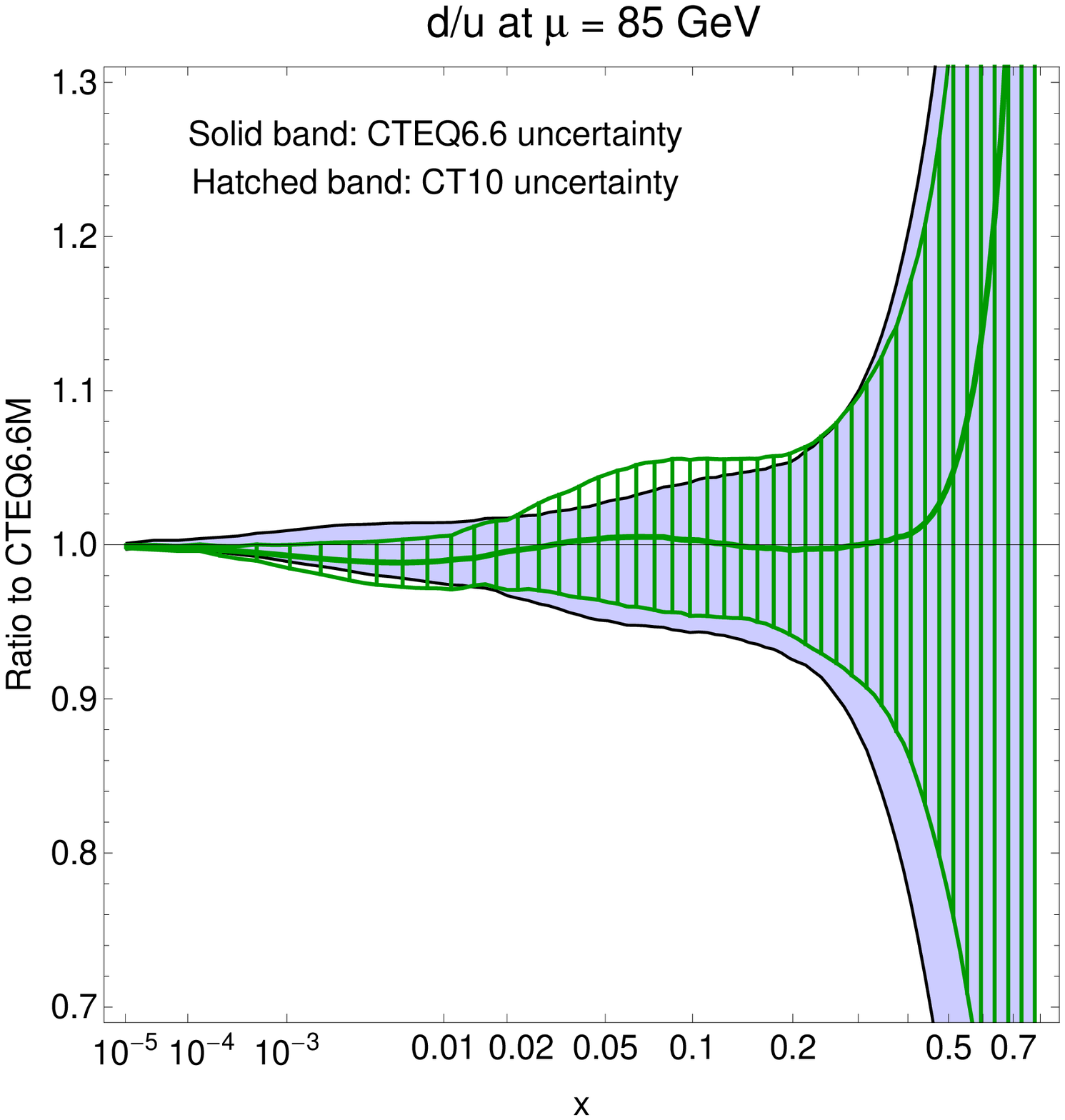}\quad\quad\includegraphics[width=0.47\textwidth]{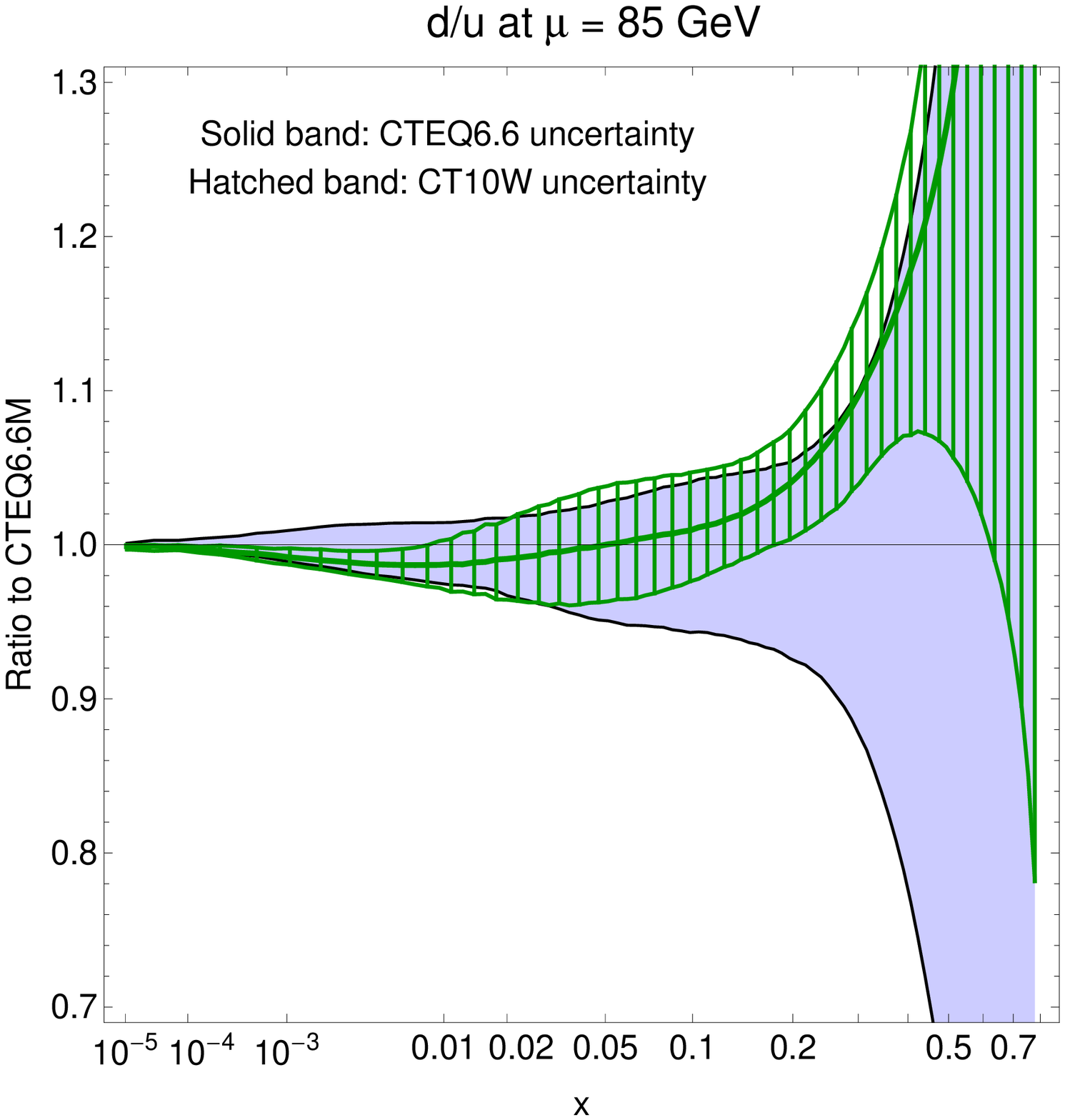}
}

{\bf Numerical predictions of the CT10W analysis.}
When included in the CTEQ global analysis, 
the CDF measurements of $A_\ell$ in Tevatron Run-1 \cite{Abe:1994rj} 
and Run-2  \cite{Acosta:2005ud}
agree well with the other data sets constraining the $d/u$ ratio, 
provided by deep inelastic scattering on proton and deuteron targets 
by the NMC \cite{Amaudruz:1992bf} and BCDMS
\cite{Benvenuti:1989rh,Benvenuti:1989fm} collaborations.  
However, the most precise Run-2 lepton $A_\ell$ data 
by the D\O~ Collaboration \cite{d0_e_asy,d0_mu_asy}
run into disagreement with the NMC and BCDMS deuteron DIS data,
and in addition, exhibit some tension among themselves.  
Because of these disagreements, 
two separate fits are produced: CT10, which does not contain the D\O~ $W$ 
electron asymmetry sets, 
and CT10W, in which they are included using weight factors 
larger than 1 to ensure an acceptable fit. We obtain
$\chi^2/Npt=91/45=2$ in the CT10W fit to the Run-2 $A_\ell$, which is
a significant improvement compared to $200/45=4.4$ obtained in
CT10. Comparison of NNLL-NLO predictions based on CT10W PDFs with the
electron charge asymmetry data is presented in Fig.~\ref{figs:eleAct10w}.

In the CT10W analysis, the inclusion of the Run-2 $A_\ell$ data increases
the slope of $d(x)/u(x)$ at $x$ between 0.1 and 0.5 and reduces its uncertainty, 
as compared to CTEQ6.6 and CT10. This is illustrated by Fig.~\ref{figs:ct10doveru}, 
which shows uncertainty bands for the $d/u$ ratio in CTEQ6.6, CT10, CT10W PDFs 
vs. the momentum fraction $x$ at scale $Q=85\mbox{ GeV }$. 

Fig.~\ref{figs:RratioZW} shows the ratios
$r_{WZ}=\sigma(pp\rightarrow W^\pm X)/\sigma(pp\rightarrow Z^0 X)$ and
$r_{W^+W^-}=\sigma(pp\rightarrow W^+ X)/\sigma(pp\rightarrow W^- X)$
of the rapidity distributions in $W^\pm$ and $Z$ boson production at the LHC,
obtained using CTEQ6.6, CT10 and CT10W PDFs 
and divided by the predictions based on the CTEQ6.6M set. 
Here, the reduction of the uncertainty bands 
in the ratio of $W^+$ to $W^-$ cross sections predicted based 
on the CT10W PDFs, as compared to CT10, is again evident. 

\shadedfigure[p]{
\caption{CT10, CT10W, and CTEQ6.6 PDF uncertainty bands for the ratios
$(d\sigma(W^\pm)/dy)/(d\sigma(Z)/dy)$ (upper two subfigures) and 
$(d\sigma(W^+)/dy)/(d\sigma(W^-)/dy)$ (lower two subfigures), at the
  LHC energies 7 and 14 TeV.}
\label{figs:RratioZW} 
\begin{centering}
\includegraphics[clip,scale=0.47]{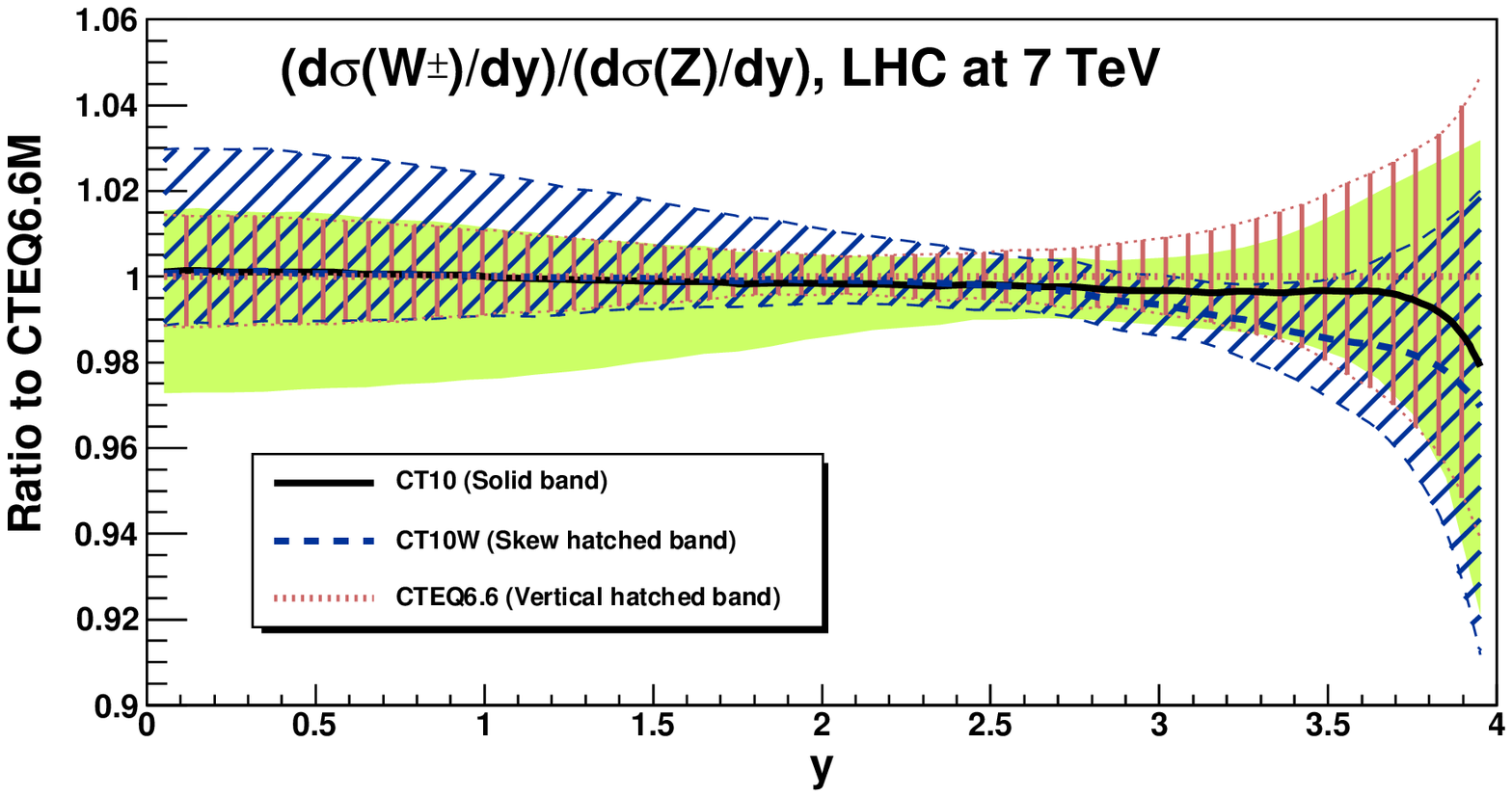}
\includegraphics[clip,scale=0.47]{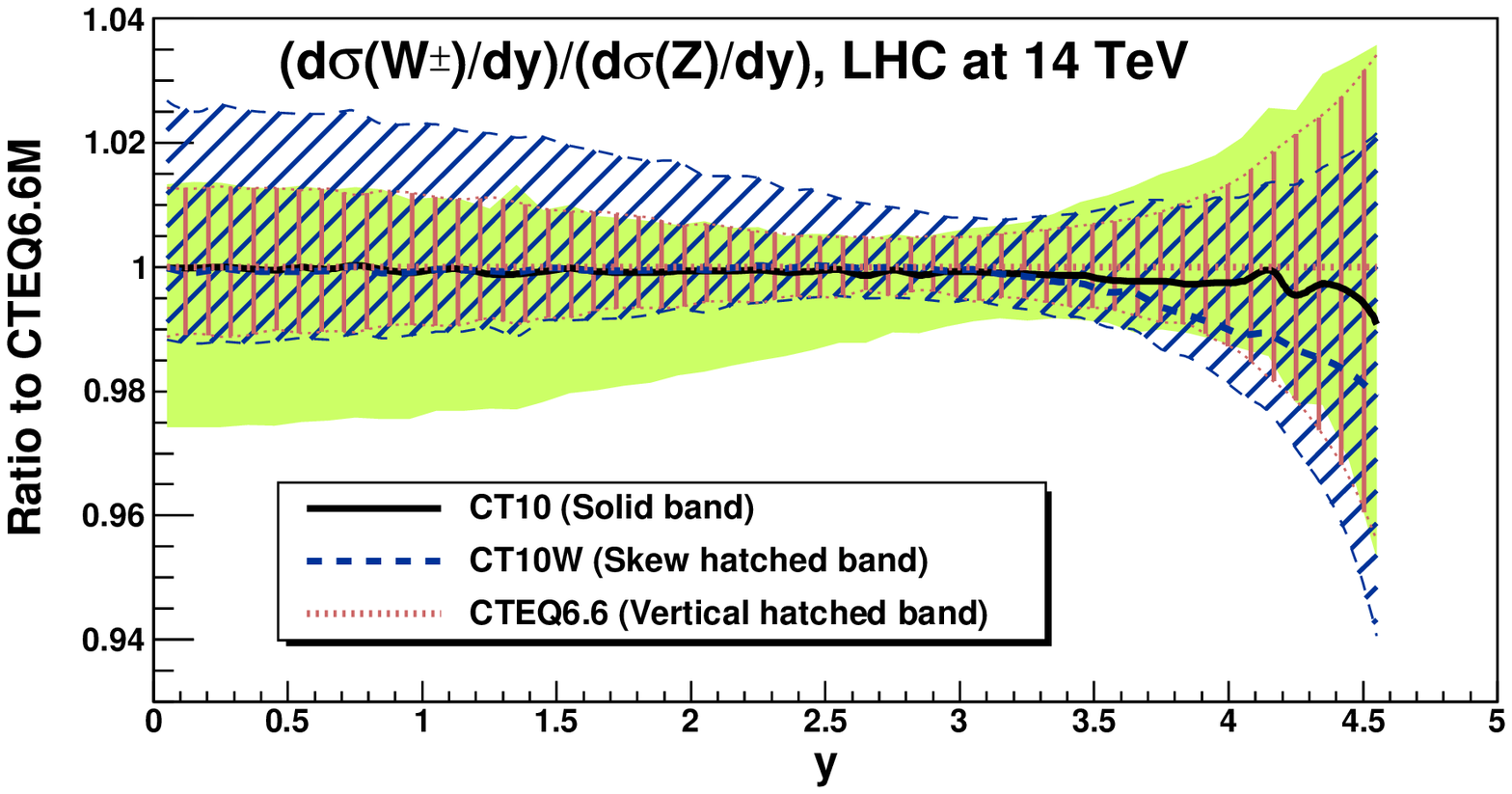}
\includegraphics[clip,scale=0.47]{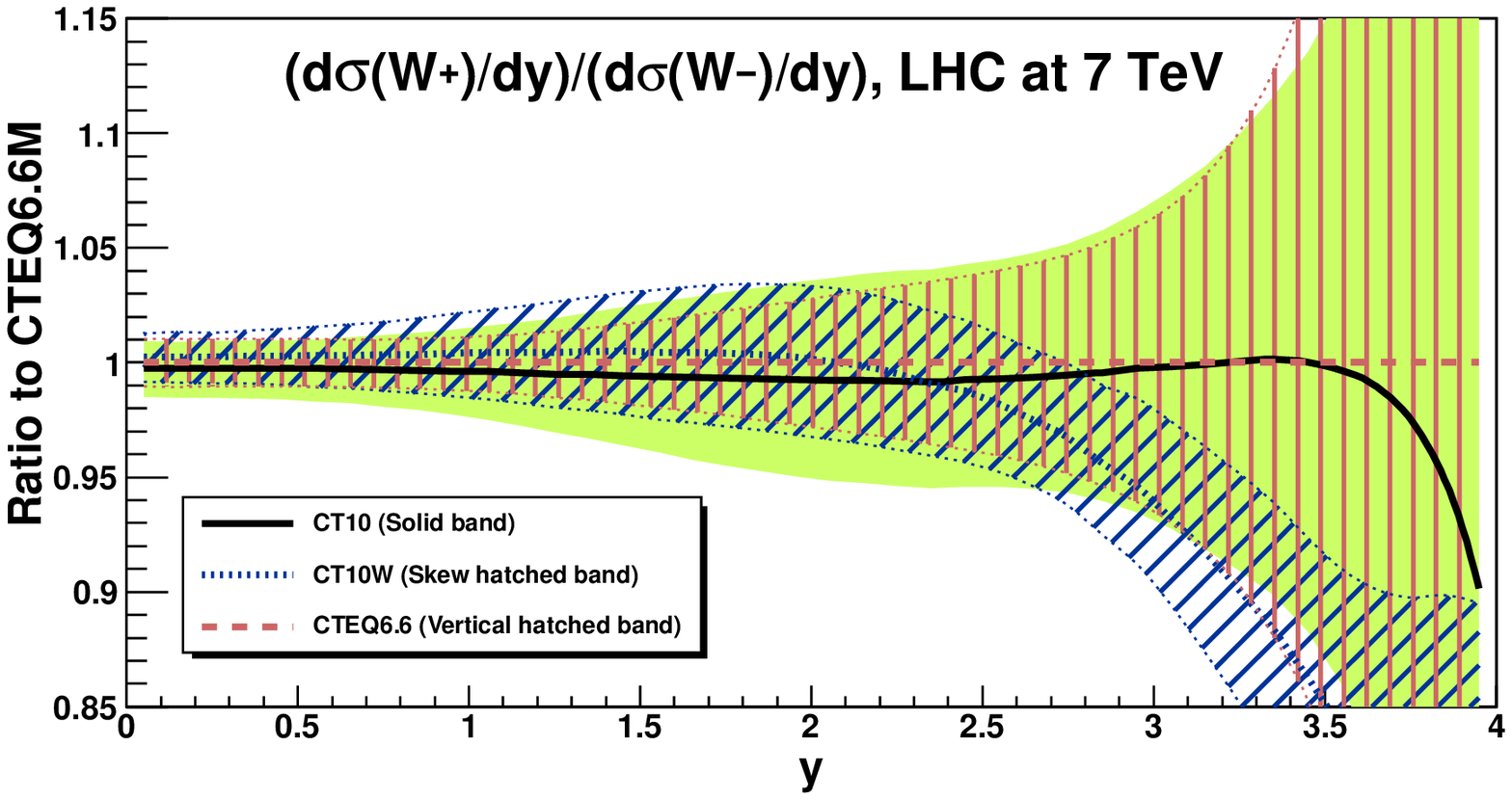}
\includegraphics[clip,scale=0.47]{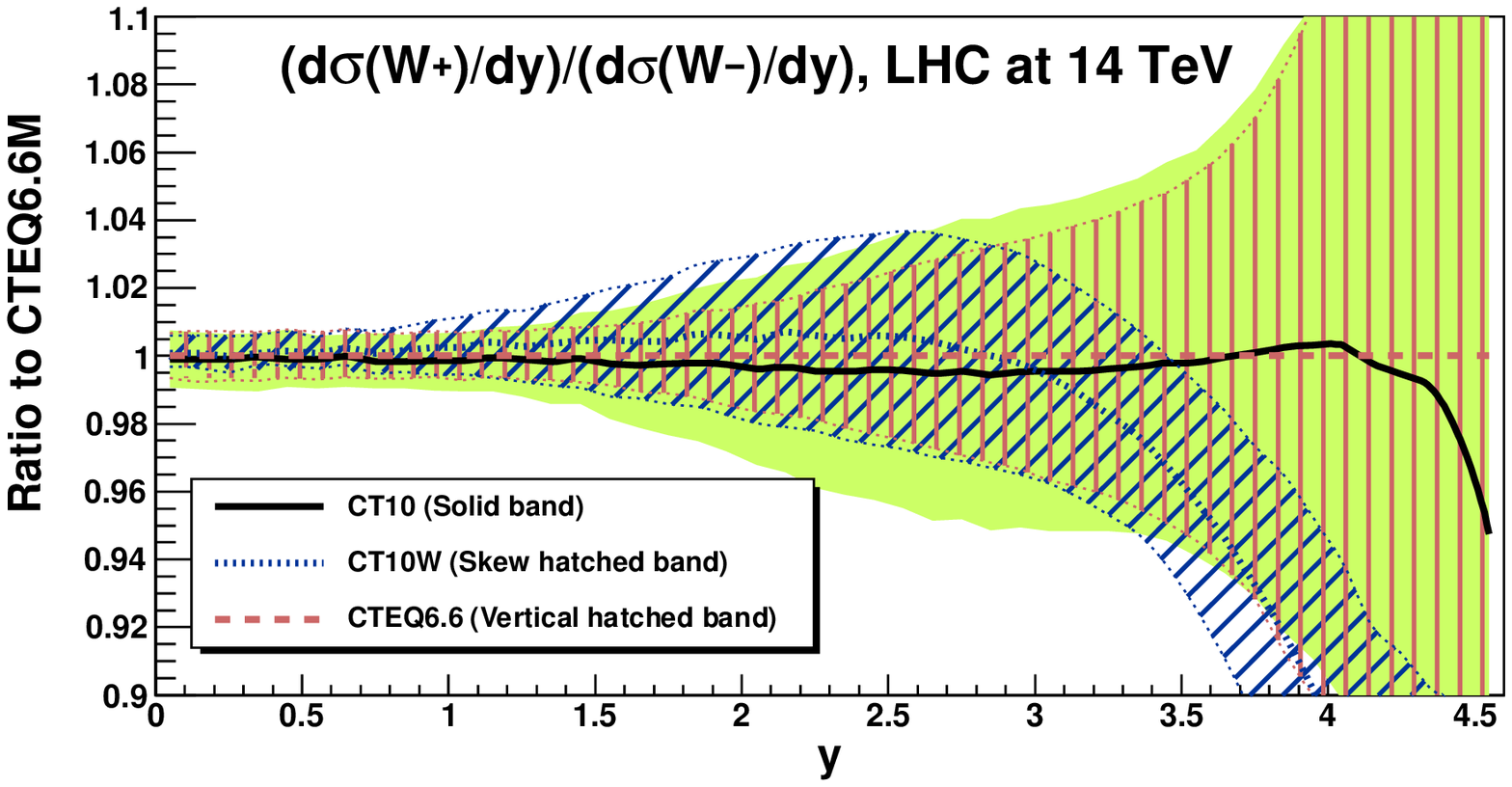} 
\par\end{centering}
}

{\bf Comparison with other fits.} 
The MSTW'08 \cite{MSTW08} and NNPDF2.0 \cite{:2010gb} groups have also
explored the impact of the D\O~ Run-2 $A_\ell$ data. 
While their conclusions are broadly compatible with ours,
the details vary. For example, in the CT10W fit, 
we find that all three $p_{T\ell}$ bins of the electron
$A_\ell$ data and the second $p_{T\ell}$ bin of the fit can be
combined with the other data sets, despite the remaining
disagreement with the deuteron DIS experiments. NNPDF, on the other
hand, finds that all bins of the muon
$A_\ell$ are compatible with the deuteron DIS and all other experiments,
but incompatible with the second and third $p_{T\ell}$ bins of the
electron $A_\ell$. NNPDF compute their NLO $A_\ell$ predictions
using the DYNNLO code \cite{Catani:2009sm}, 
which can deviate from the resummed NNLL+NLO predictions by ResBos used by CTEQ, as
discussed above. 

The resolution of the Run-2 $A_\ell$ puzzle thus
seems to require consistent implementation of perturbative QCD
calculations both in experimental analyses and PDF fits, including
small-$Q_T$ resummed contributions, when constraints on the lepton
$p_{T\ell}$ are imposed.  For greater precision it may be preferable 
to perform the small-$Q_T$ resummation for 
every component of the angular distribution of the decay lepton \cite{Berger:2007jw}, 
in addition to the resummation for two dominant angular components 
that is currently implemented in ResBos.  Resummation of the full dependence 
on the polar and azimuthal angles $\theta$ and $\phi$ of the lepton in the vector
boson rest frame may be important especially in situations in which the experimental
coverage is not uniform in all directions, or when there are gaps in
the coverage.  Normally the angular coverage varies with rapidity, notably
at large rapidity, and this variation may well affect the
lepton asymmetry, with different consequences in different experiments. 

\clearpage \newpage

\subsection{CT10 predictions for collider observables}
Fig.~\ref{figs:XSECct10w} compares the NLO total cross sections, obtained  
using CT10 and CT10W PDFs, to those obtained using CTEQ6.6 PDFs, for some selected  
processes at the Tevatron Run-2 and the LHC at $\sqrt{s}=7\mbox{ TeV}.$
For most of the cross sections, CT10
and CT10W sets provide similar predictions and uncertainties, which 
are also in good agreement with those from CTEQ6.6 ({\it i.e.}, well within the PDF
uncertainty band). 
At the LHC, the PDF uncertainties in 
CT10 and CT10W predictions for some processes are larger than
those in the counterpart 
CTEQ6.6 predictions, reflecting the changes in the framework
of the fit discussed in the next paragraph. At the Tevatron, 
the CT10(W) PDF uncertainties 
tend to be about the same as those for CTEQ6.6, with a notable
exception of $t\bar t$ production cross sections, which have
a smaller PDF uncertainty with the CT10W set, because of 
stricter constraints on the up- and down-quark PDFs at the relevant $x$ values. 
\shadedfigure[p]{
\caption{Ratios of NLO total cross sections obtained using CT10 and
  CT10W to those using CTEQ6.6M PDFs, in various scattering
  processes at the Tevatron Run-II and LHC.}
\label{figs:XSECct10w} 
\begin{centering}
\includegraphics[width=0.49\textwidth,height=0.45\textheight]{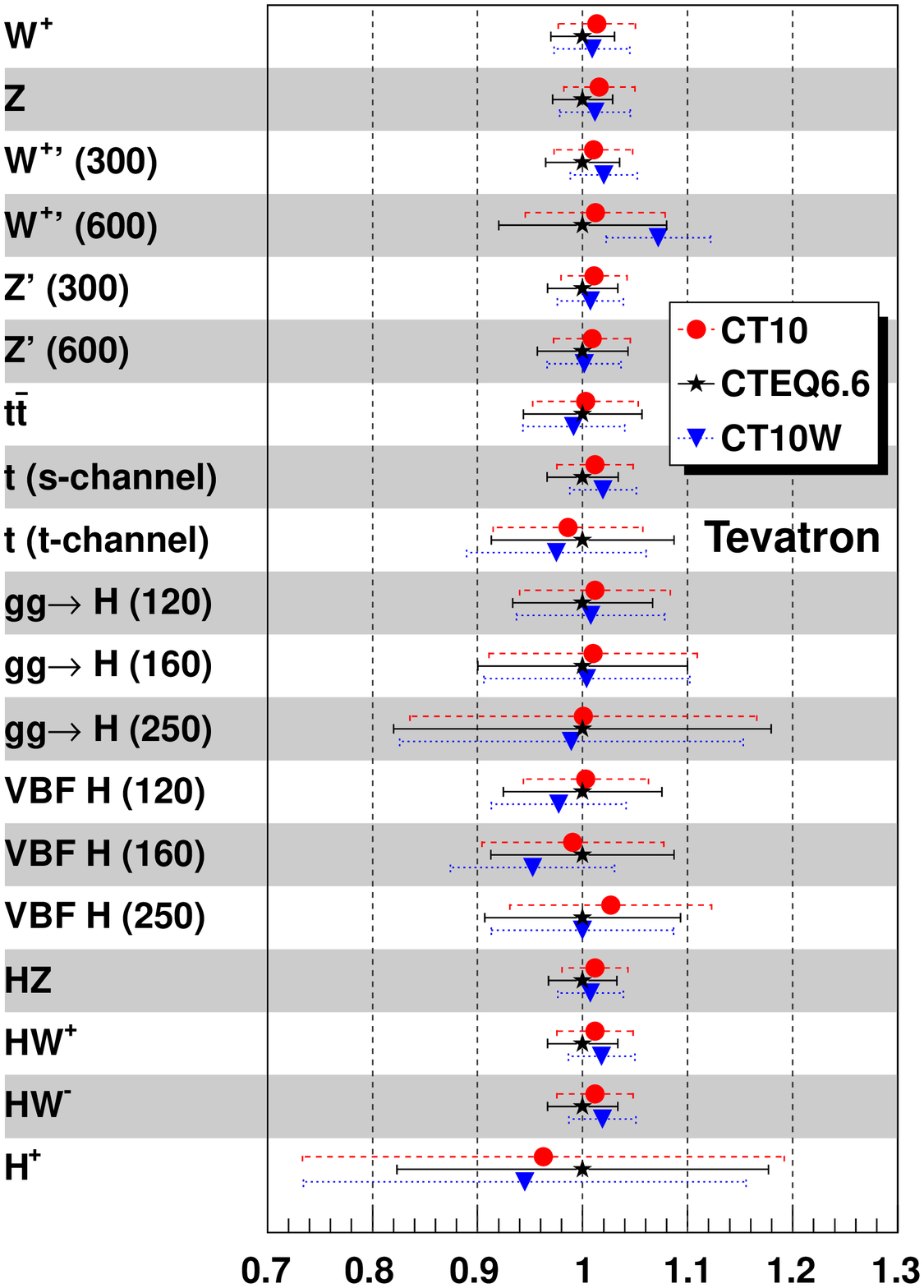}
\includegraphics[width=0.49\textwidth,height=0.45\textheight]{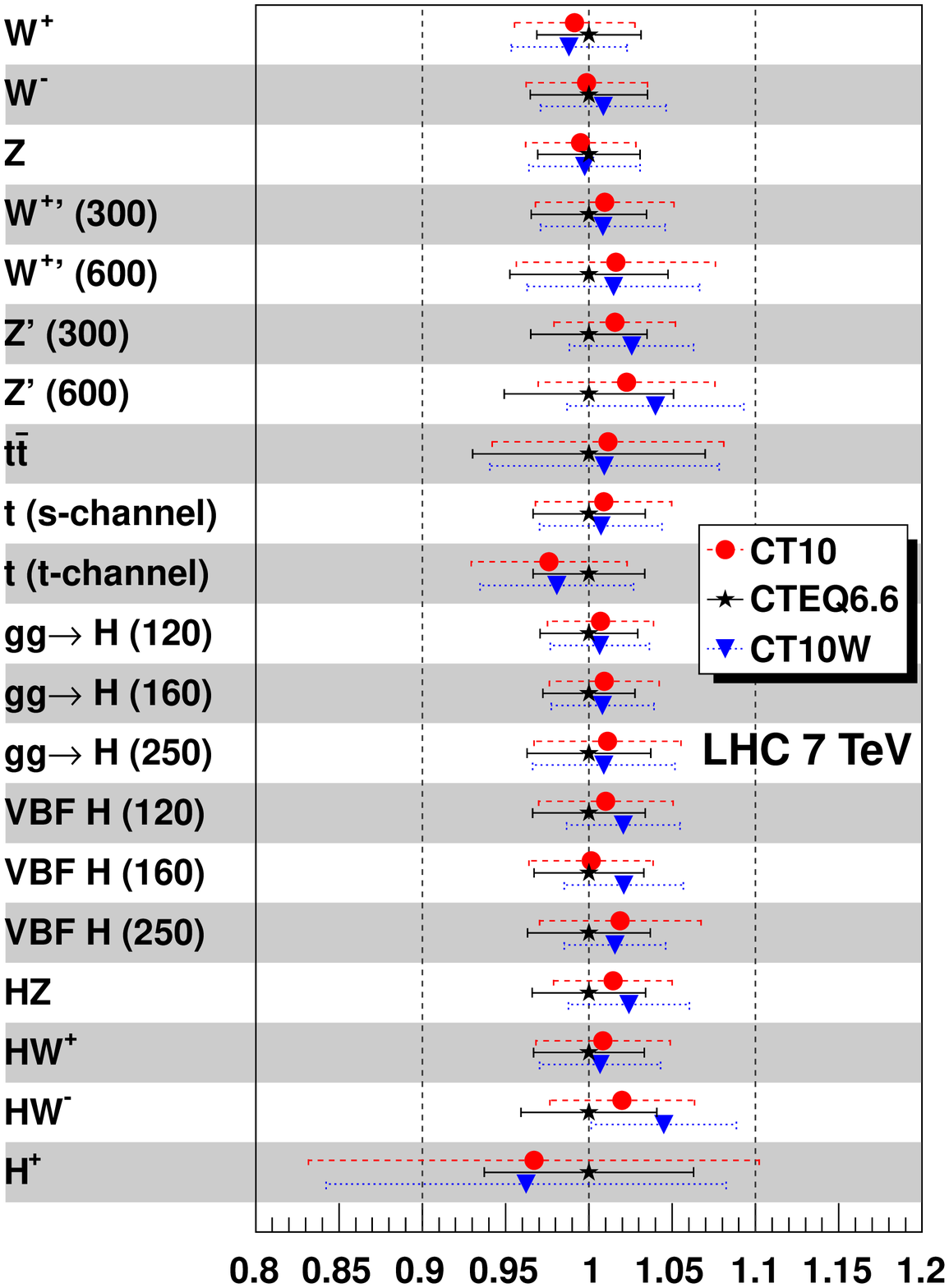}
\includegraphics[width=0.49\textwidth,height=0.45\textheight]{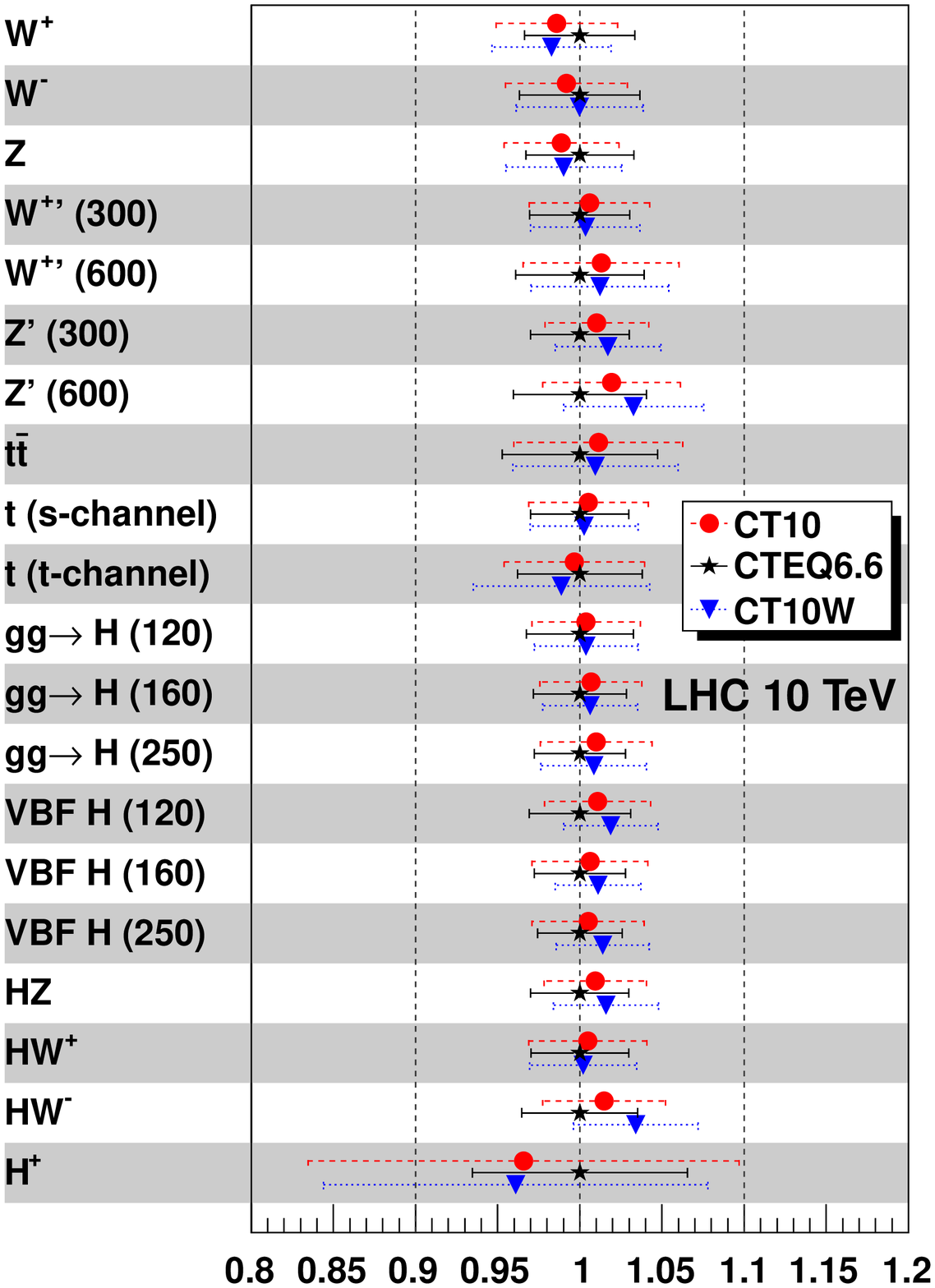}
\includegraphics[width=0.49\textwidth,height=0.45\textheight]{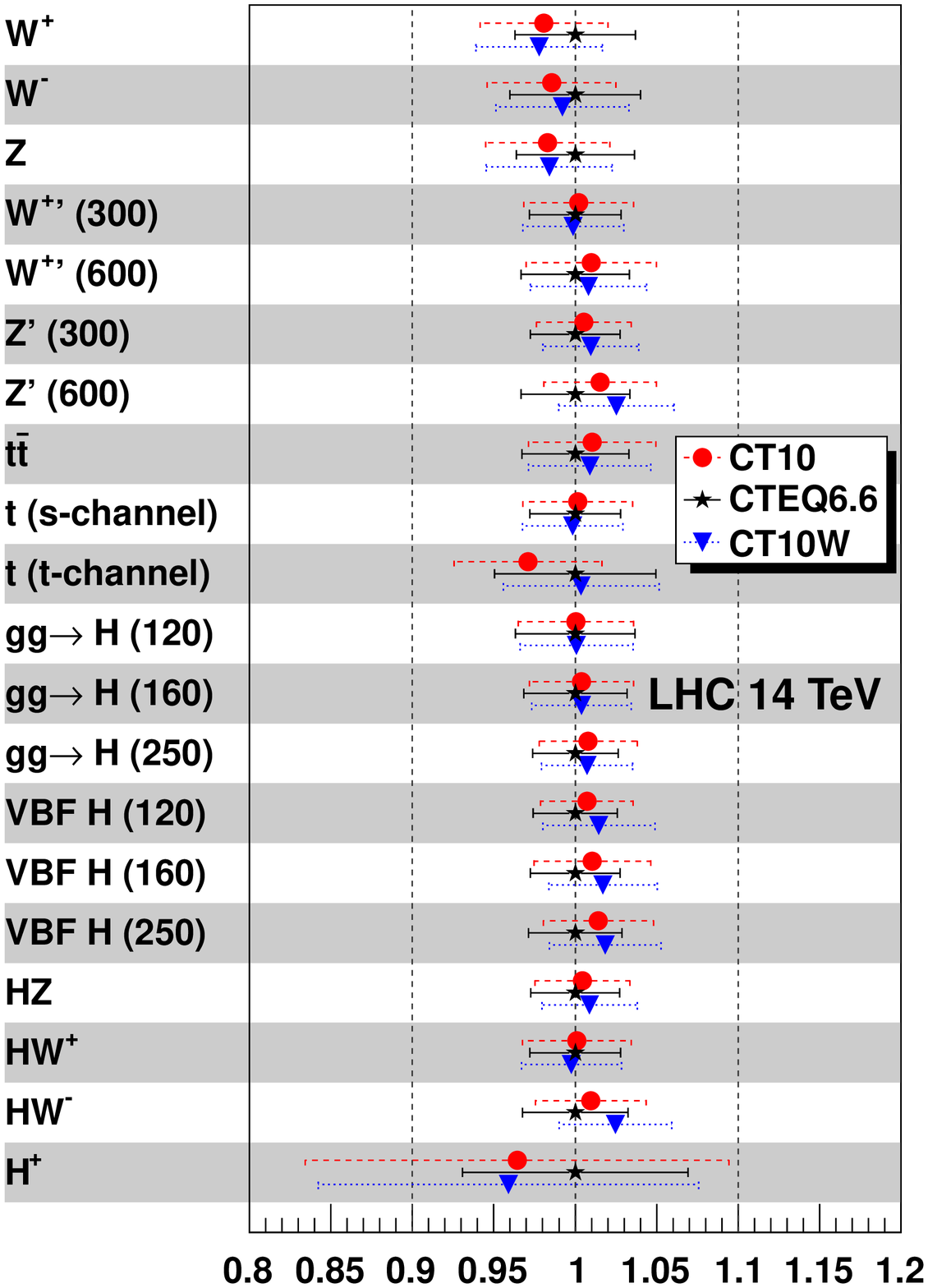} 
\par\end{centering}
}

\shadedfigure[p]{
\caption{Comparison of \protect{D\O~} Run-II data for dijet invariant mass distributions  \cite{d0-di-jet} with NLO theoretical predictions and their PDF uncertainties for CTEQ6.6 (black), CT10 (red) and CT10W (blue) PDFs. The cross sections are normalized to theoretical predictions based 
on the best-fit CT10 set, designated as CT10.00.}
\label{figs:DIJET6} 
\includegraphics[width=0.49\textwidth]{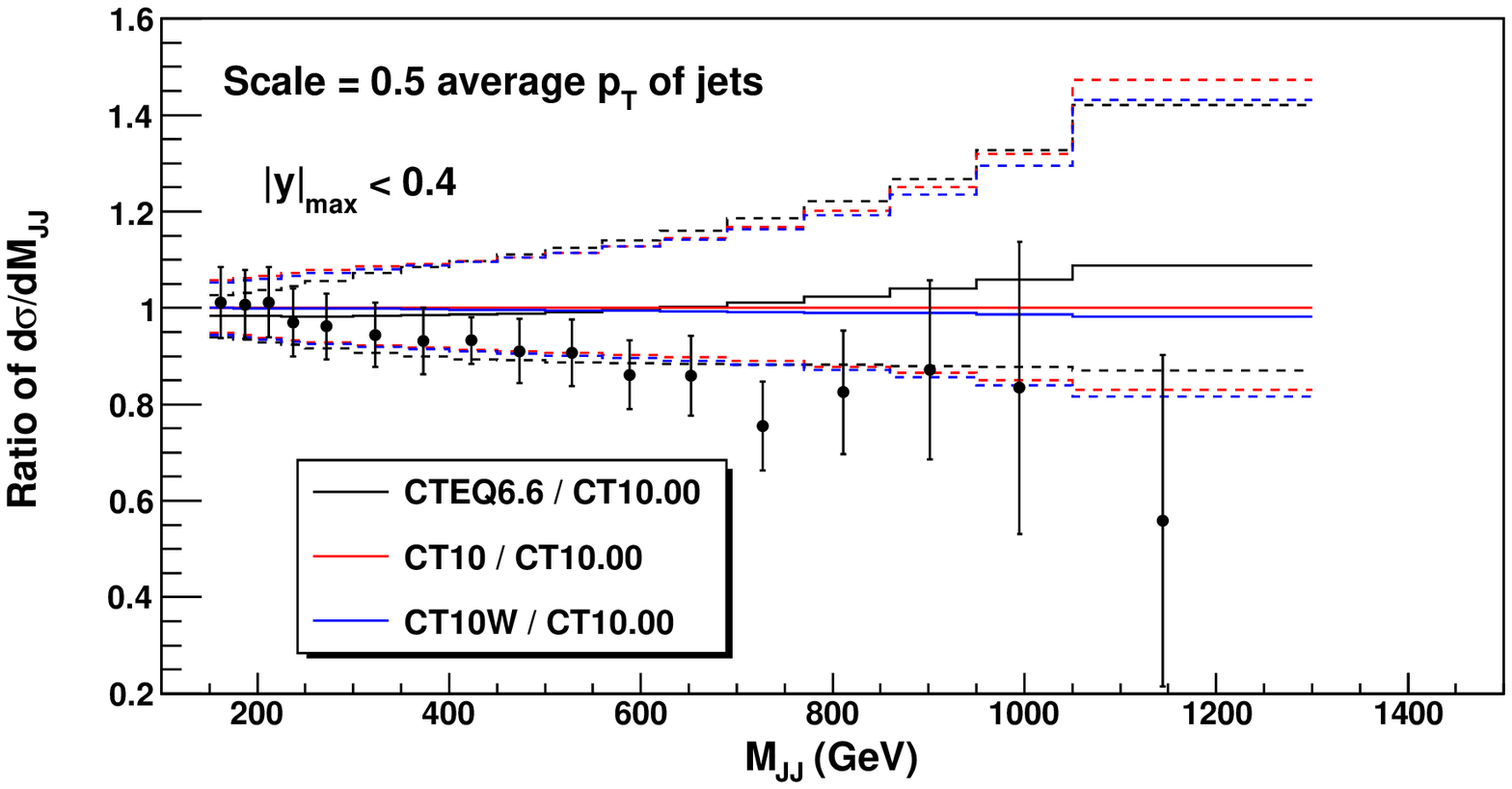}
\includegraphics[width=0.49\textwidth]{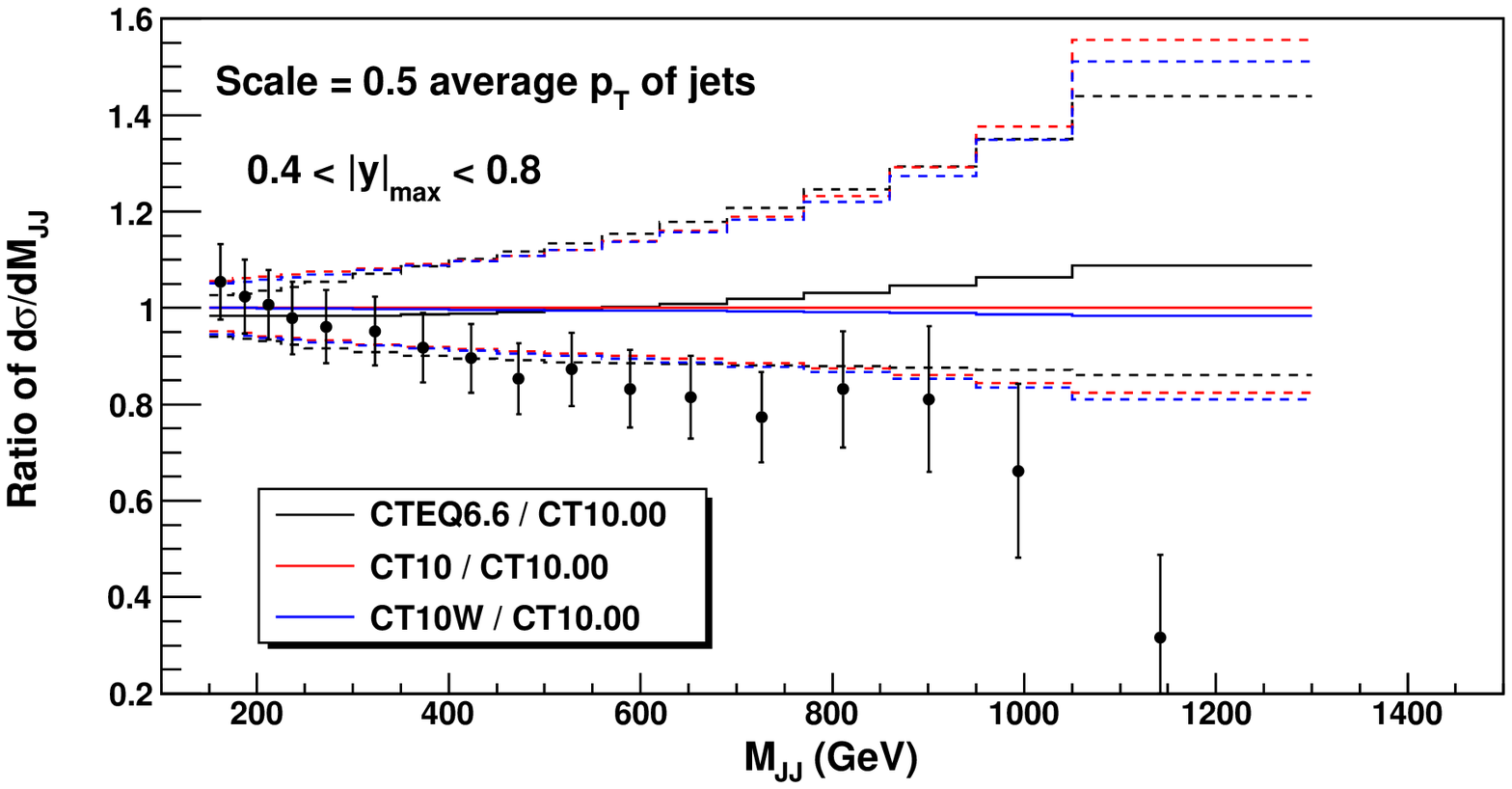}
\includegraphics[width=0.49\textwidth]{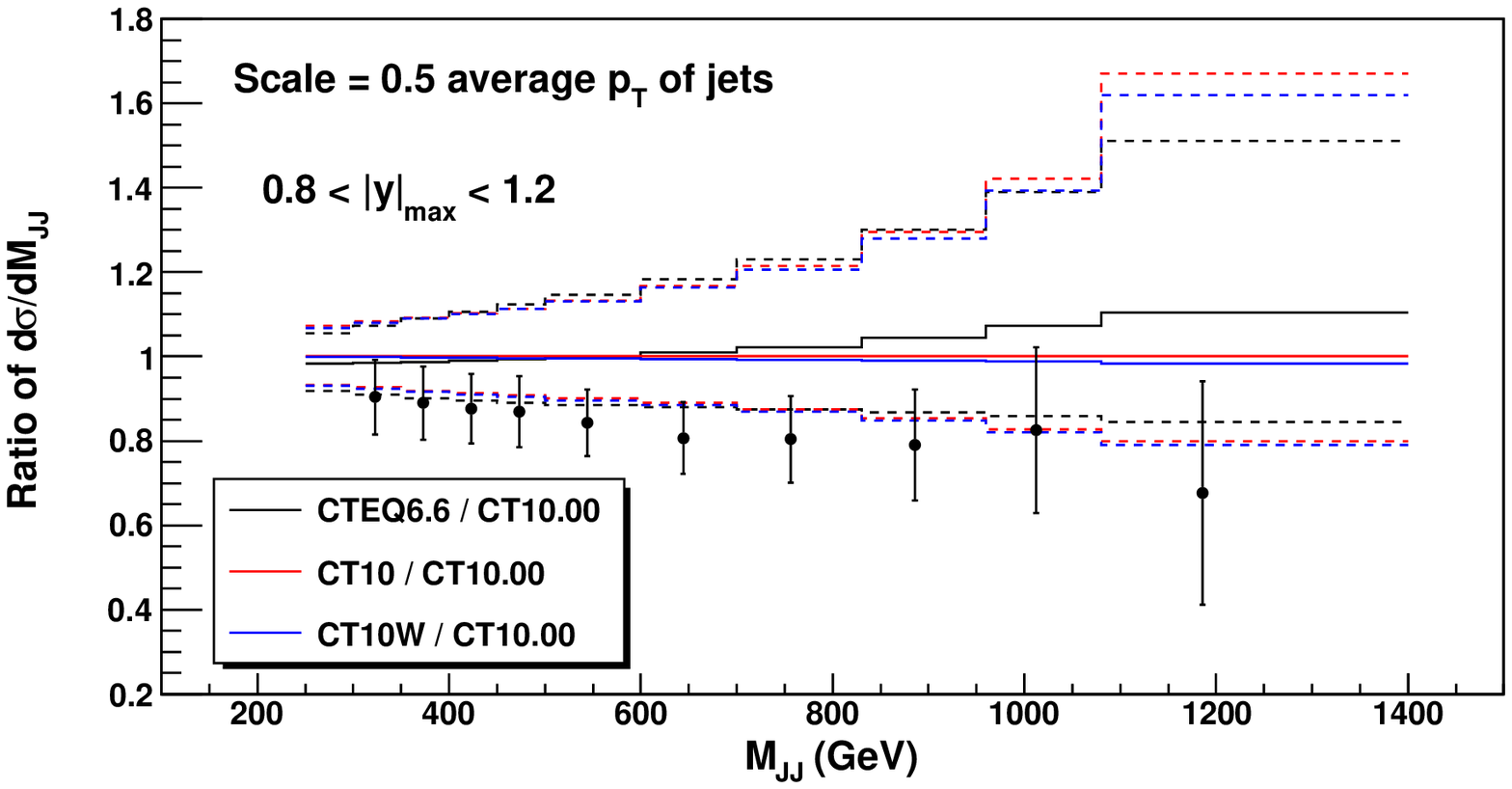}
\includegraphics[width=0.49\textwidth]{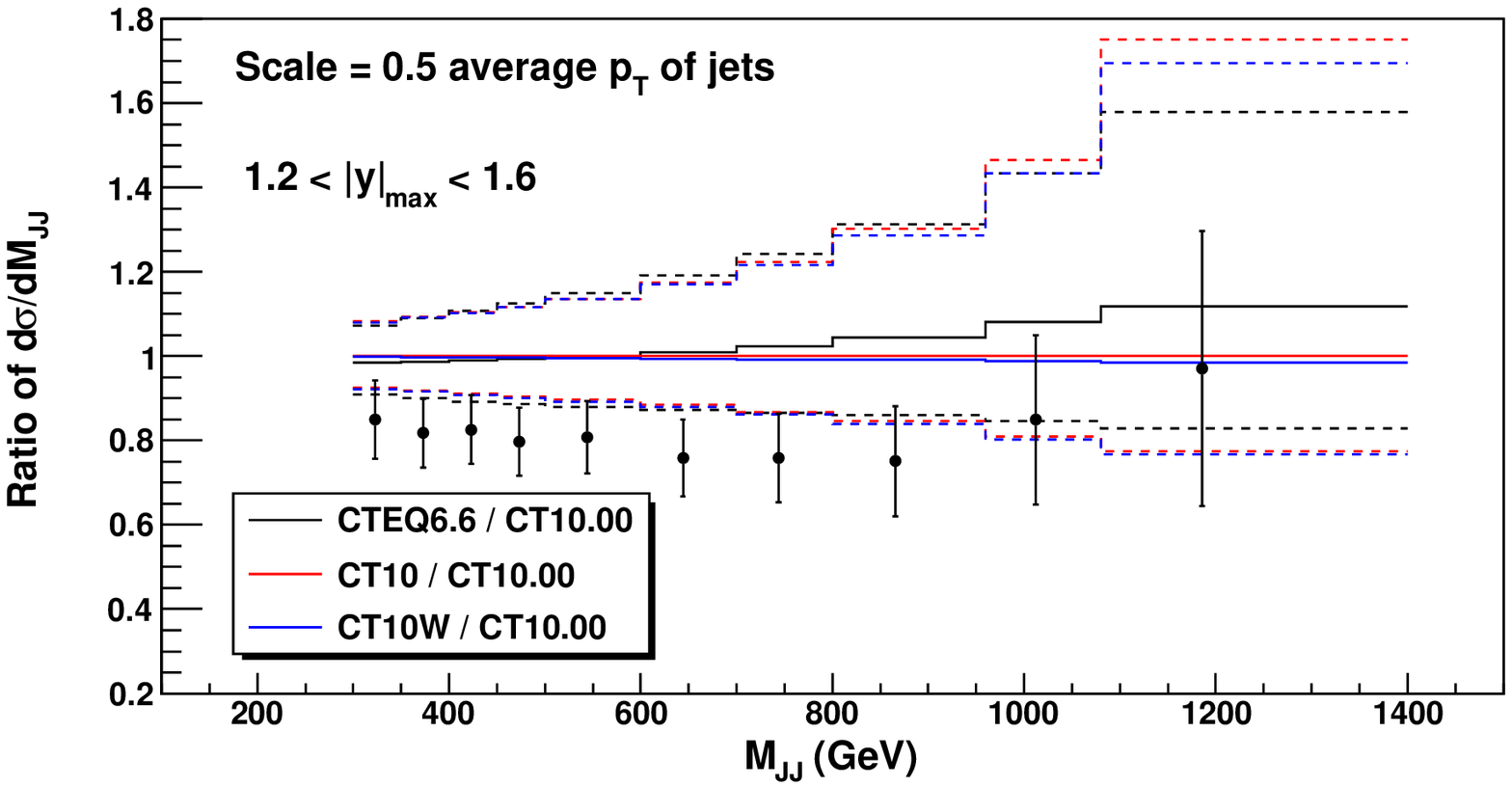}
\includegraphics[width=0.49\textwidth]{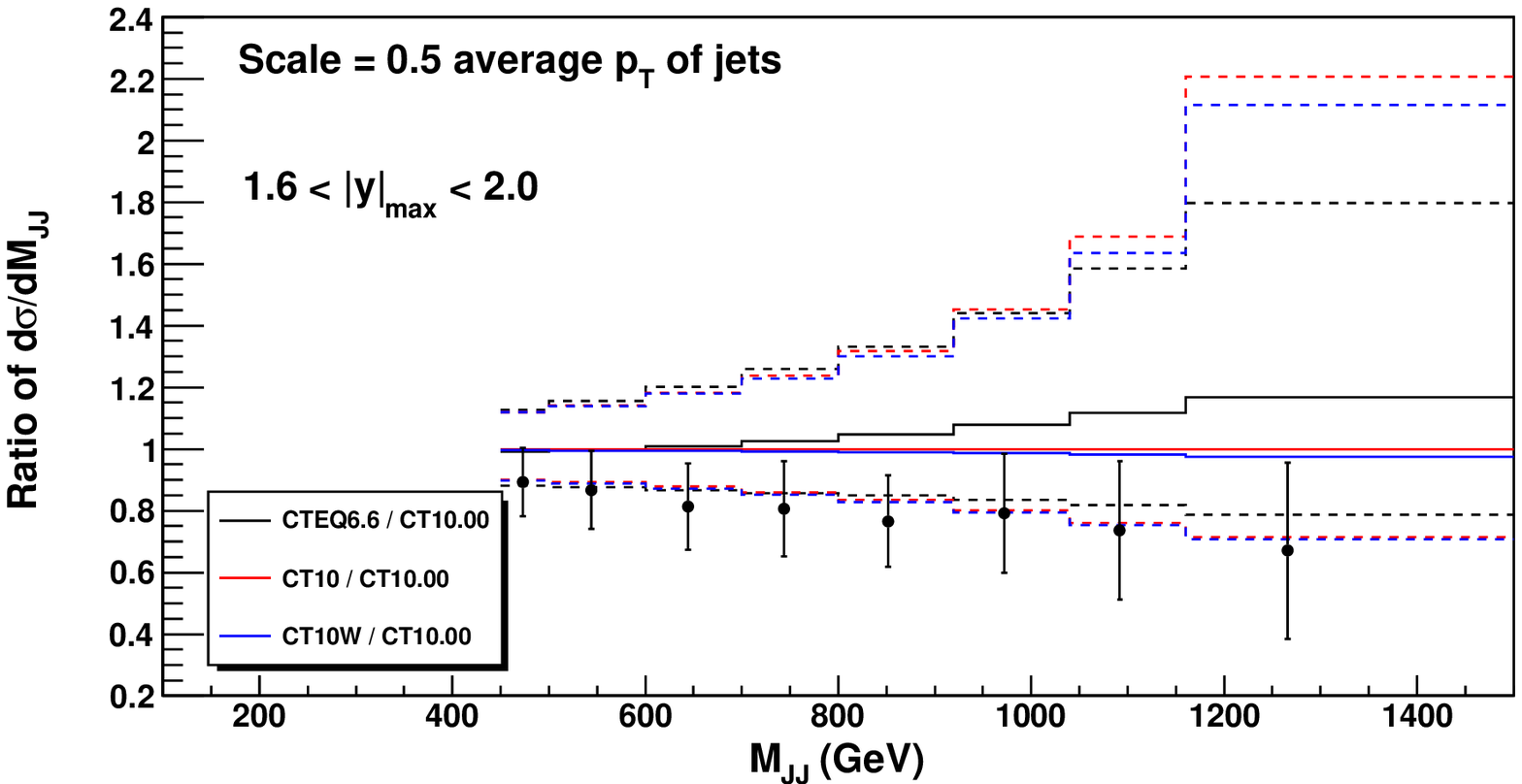}
\includegraphics[width=0.49\textwidth]{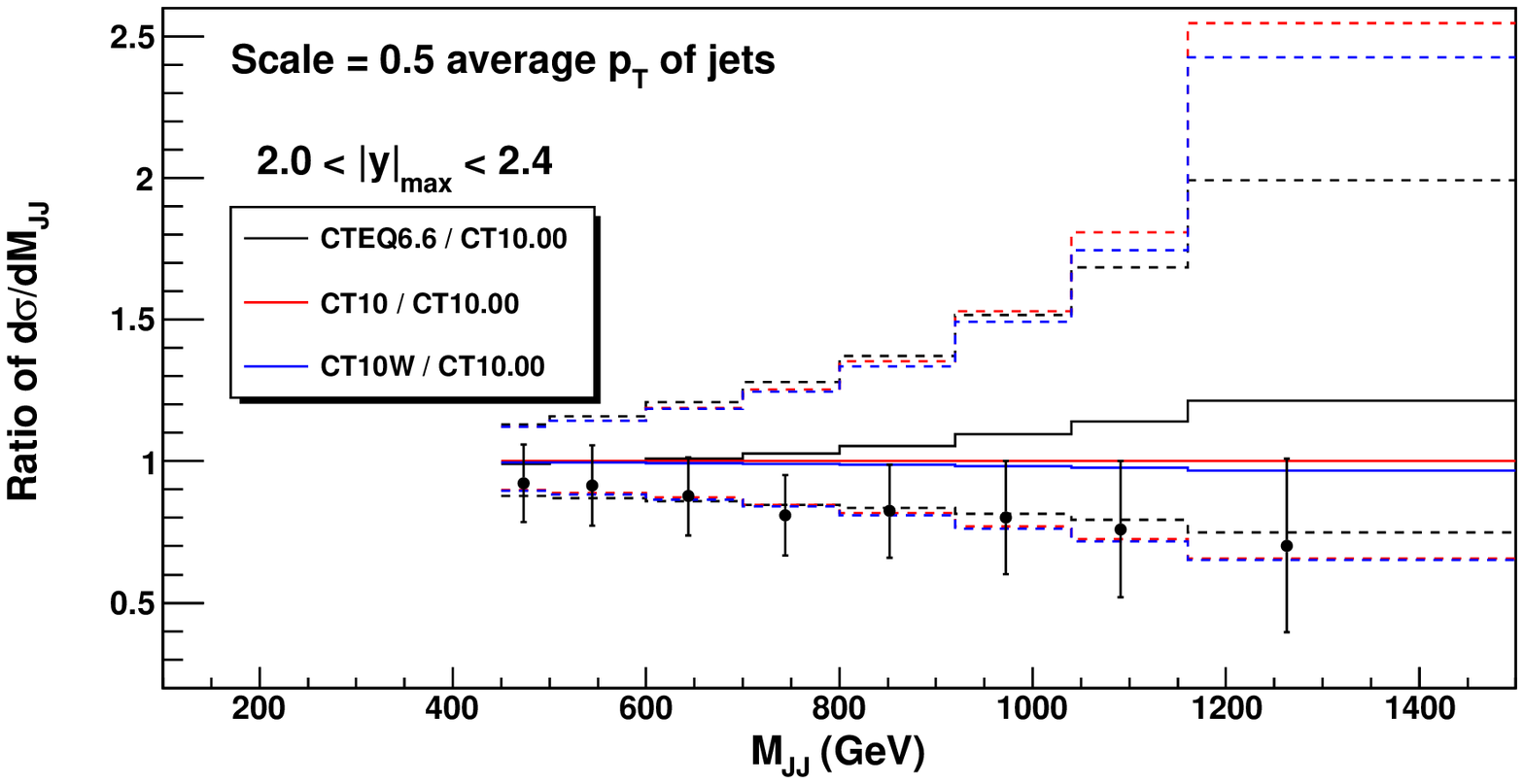}
}

\subsection{New PDF parametrizations; advancements in statistical analysis}
The CT10 global analysis implements several new features which were
not available in the previous studies. The systematic uncertainty associated with the
overall normalization factor in each of the data sets is handled in the same 
manner that all other systematic error parameters are handled. The
best-fit values of the normalizations are found algebraically, and
their variations are included in the final estimate of the PDF
uncertainties. More flexible parametrizations are assumed for the
gluon, $d$-quark, and strange quark PDFs at the initial scale
$1.3\mbox{ GeV}$, to reduce biases in predictions in kinematical regions
where the constraints from the data are weak. Finally, a new
statistical procedure is introduced to guarantee the
agreement of the fits at 90\% C.L. with all included experiments, for
any PDF eigenvector set produced by the error analysis. This is
realized by adding  an extra contribution to the total $\chi^{2}$, which guarantees  
the quality of fit to each individual data set and halts 
the displacement along any eigenvector early, if necessary,
to prevent one or more individual data sets from being badly
described. The old procedure to enforce the 90\% C.L. agreement with
all experiments in the CTEQ6 family of fits, by artificially
increasing statistical weights of $\chi^2$ contributions from those
experiments that may be fitted poorly by some PDF eigenvector sets, is
phased out by this more efficient method. 
As a result of these changes, the CT10/CT10W PDF uncertainty may be smaller
or larger than the CTEQ6.6 uncertainty, depending on whether the
improved constraints from the data outweigh the increased uncertainty
due to the relaxed PDF parametrizations and variations of normalizations during
the determination of PDF eigenvector sets.

\subsection{Dijet invariant mass distributions}
Fig.\ref{figs:DIJET6} compares NLO predictions
based on CTEQ6.6, CT10, and CT10W PDFs  with the data on the dijet invariant mass
distribution $d\sigma/dM_{jj}$ reported recently by the D\O~
Collaboration \cite{d0-di-jet}. These data are not included in the
CT10 fit, but they are sensitive to the same scattering subprocesses,
and include the same events, as the single-inclusive jet data
constraining the gluon PDF in the CT10 fit. 
The cross sections are normalized to
the theory prediction based on the central CT10 PDF set.
The statistical and systematic errors of the
D\O~ data are added in quadrature. 
The renormalization and factorization scales in the theoretical
predictions are set equal to a half of the average $p_T$ of the jets,
$\langle p_T\rangle/2$, consistent with the scale used in the single-jet cross
sections when determining the gluon PDFs. 
With this choice of the scale,  
all three PDF sets agree with most of the data points within the PDF
uncertainty. There appears to be a systematic excess of theory over the
data, but its magnitude strongly depends on the assumed factorization
scale. For example, it can be much worse if a different scale is
taken, such as $\mu = \langle p_T \rangle$ assumed in Fig. 2 
of the D\O~ paper \cite{d0-di-jet}. We conclude that the CT10 PDFs are
reasonably compatible with the D\O~ dijet data within the present
theoretical uncertainties, 
although an overall systematic shift, of order of the systematic
shifts observed in the CT09 study of the single-inclusive jet
distributions \cite{Pumplin:2009nk}, would further improve the
agreement, once the full correlated systematic errors 
of the dijet data become available. 

\subsection{Uncertainty due to $\alpha_s$ in CTEQ6.6 and CT10 PDF analyses}
Many calculations for collider processes 
({\it e.g.}, production of $t\bar t$ pairs and
Standard Model Higgs bosons) require to evaluate two
leading theoretical 
uncertainties, due to the choice of the PDF parametrization at 
the initial scale, and the value of the strong coupling constant 
$\alpha_s(M_Z)$ assumed in the hard cross sections and the
PDFs. These uncertainties can be comparable in size, and their 
interplay, or correlation, may be
important.  In Ref.~\cite{Lai:2010nw}, 
we examine the $\alpha_s$ dependence of CTEQ6.6 PDFs
\cite{Nadolsky:2008zw} and show
how the PDF-$\alpha_s$ correlations are adequately captured by a 
simple calculation, without resorting to more elaborate methods proposed 
in other studies \cite{Martin:2009bu}.

At the beginning of the PDF fit, one decides which data 
determine the $\alpha_s(M_Z)$ value and its uncertainty. 
CTEQ {\em best-fit} PDFs and
their parametrization uncertainties are usually 
determined for a constant value 
of $\alpha_{s}(M_{Z})$ that is close to its latest world-average central
value; e.g.,  $\alpha_{s}(M_{Z})=0.118\pm 0.002$ 
assumed in Ref.~\cite{Lai:2010nw}. This {\it input} value of
$\alpha_s(M_Z)$ can be viewed as an additional data point that
summarizes world constraints on $\alpha_s$, mostly determined by
precise experiments that are not included in the global fit (notably, 
LEP event shapes and $\tau$ and quarkonium decays). 
In Ref.~\cite{Lai:2010nw}, we explore a more general procedure, in which the
world-average data point on $\alpha_s(M_Z)$ is included in the fit 
in addition to the usual hadronic scattering data. A 
theoretical parameter for $\alpha_s(M_Z)$ is varied in this fit; 
its {\it output} value and uncertainty are determined by all input data.
We find that the output value of $\alpha_s(M_Z)=0.118\pm 0.0019$ 
obtained in this way essentially
coincides with its input value $\alpha_s(M_Z)=0.118\pm 0.002$. 
If the input value is not included, the output uncertainty 
on $\alpha_s$ is increased significantly, to $\alpha_s(M_Z)=0.118\pm 0.005$.
This indicates that the hadronic scattering data included 
in the fit imposes significantly weaker
constraints on $\alpha_s$ than the other experiments contributing to the
world-average value.

The Hessian PDF eigenvector sets returned by such floating-$\alpha_s$ fit 
can be used to estimate  the correlation between the PDF parameters 
and $\alpha_s(M_Z)$. However, each of these eigenvector sets is inconveniently 
associated with its own value of $\alpha_s(M_Z)$. Instead, 
one can apply a simpler procedure, in which all eigenvector sets 
except two are determined for the best-fit value of
$\alpha_s(M_Z)$. These eigenvector sets provide the usual
PDF uncertainty for a fixed $\alpha_s(M_Z)$. 
Separately, the uncertainty in the PDFs induced by 
the uncertainty in $\alpha_{s}(M_{Z})$ is assessed, 
by producing two alternative PDF fits for the $\alpha_{s}(M_{Z})$ 
values at the lower and upper ends of the $\alpha_s$ uncertainty interval 
({\it i.e.}, $\alpha_s(M_Z)=0.116$ and $0.120$).
These PDF and $\alpha_s$ uncertainties are then added in quadrature 
to obtain the total uncertainty.

This procedure is valid both
formally and numerically. It is based on a theorem that is applicable
within the quadratic approximation for the log-likelihood 
function $\chi^2$ in the vicinity of the best fit. The proof of the
theorem, as well as a numerical demonstration of the equivalence of
the addition in quadrature to the full estimation of the
PDF+$\alpha_s$ uncertainty based on the Hessian method, are given in
Ref.~\cite{Lai:2010nw}. The series of best-fit PDFs for
$\alpha_s(M_Z)$ values in the interval 0.113-0.123, needed to evaluate
the combined PDF+$\alpha_s$ uncertainty in any application, 
are made available both for CTEQ6.6 and CT10 PDFs.

\shadedfigure[tb]{
\caption{\small{Left: Ratio of $g(x,Q)$ in SUSY fits with a fixed $\alpha_s(M_Z)=0.118$ and CT10 fit. 
Right: $\Delta\chi^2$ in 2004 and 2010 SUSY fits vs. gluino mass.}} 
\label{res}
\begin{center}
\includegraphics[width=5cm, angle=-90]{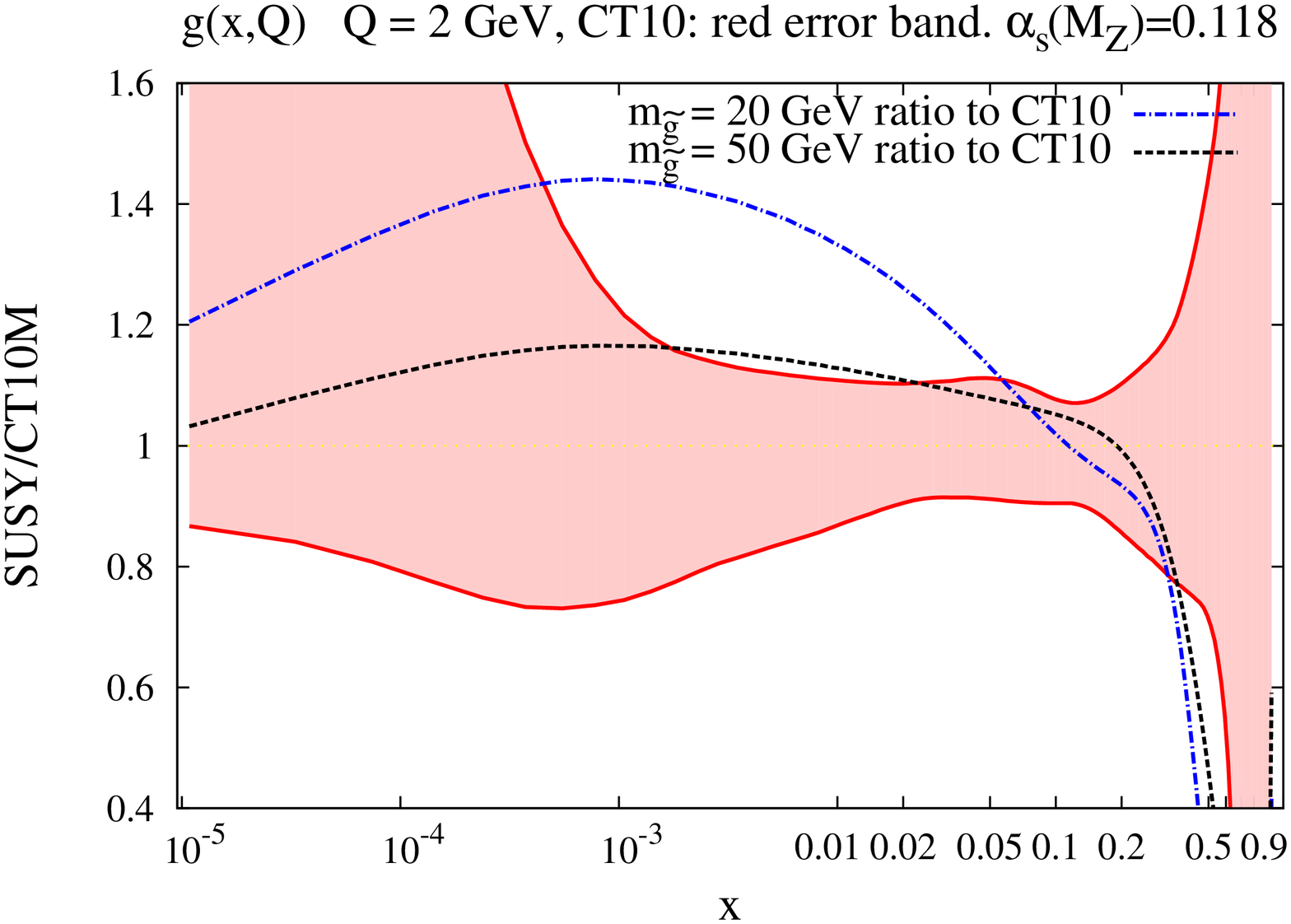}
\includegraphics[width=5cm, angle=-90]{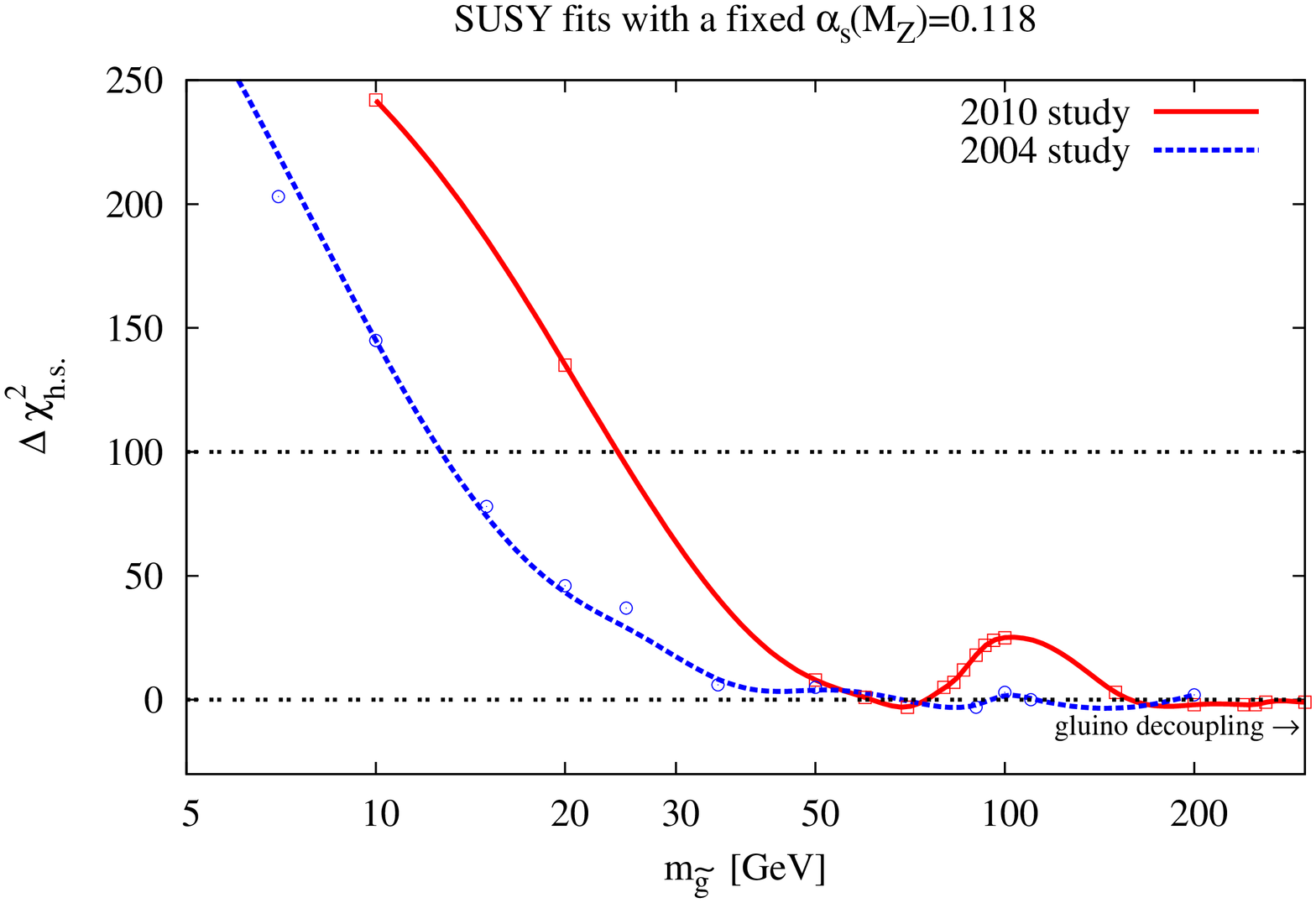}
\end{center}
}

\subsection{Constraints on new physics from a global PDF analysis} 
Besides providing the PDFs and their uncertainties, 
the global QCD analysis can establish 
bounds on masses of hypothetical particles beyond the
standard model (BSM), for example, relatively light color-octet
Majorana fermions that contribute to strong interaction processes with
the same coupling strength as the gluons. Gluinos of a supersymmetric (SUSY)
origin serve as an example of such fermions; but other models can introduce
them as well. Constraints on the ``gluinos'' are
often imposed in the context of a specific BSM model for their
production and decay, which helps to rule out ``gluinos'' 
with masses up to a few hundred GeV. But, if no model-specific
assumptions are imposed, much lighter gluinos are
allowed: as light as $\approx 12-15$ GeV 
according to the 2004 global fit based on the CTEQ6 set data
\cite{Berger:2004mj}, or 6-51 GeV 
according to the NLO/resummation analyses of $e^+e^-$ hadroproduction 
at LEP \cite{Janot:2003cr,Heister:2003hc,Kaplan:2008pt}. 
Note that the limit based on the global fit is not affected by
(potentially important) theoretical uncertainties in the LEP analyses, 
associated with nonperturbative and matching effects in the
resummation techniques that they employ.

In Ref.~\cite{Berger:2010rj}, we improve the earlier limits 
\cite{Berger:2004mj} on relatively light gluinos 
based on an extended CT10 fit with added
gluino scattering contributions. In this study, an independent PDF 
describing gluinos, and one-loop splitting functions
describing interactions of gluinos with quarks and gluons, are
introduced in the DGLAP equation. Two-loop 
gluino contributions are included in the renormalization group
equation for the running of $\alpha_s$. Cross sections for 
single-inclusive jet production are modified to include 
hard matrix elements for $2\rightarrow 2$ processes involving gluinos,
with full dependence on the gluino mass evaluated in the general-mass
factorization scheme. 
These modifications capture the essential dependence on gluinos 
in inclusive processes studied in the CT10
fits. Generally, the gluino contributions to inclusive observables are
small, so that they can be evaluated at the one-loop level to achieve
the same accuracy as the SM contributions evaluated at two
loops. Squarks and other BSM particles are assumed
to be heavier than a few hundred GeV and not included.

Constraints on the gluino mass $m_{\tilde g}$ 
depend strongly on the value of
$\alpha_s(M_Z)$  \cite{Berger:2004mj}. To reproduce the existing
limits on $\alpha_s(Q)$, we
introduce two data points at $Q=5$ GeV and $M_Z$, representing a
combination of measurements at low $Q$ and $Q\approx M_Z$,
respectively.
The low-$Q$ bound on $\alpha_s$ has been obtained by combining 
measurements of $\alpha_s(M_Z)$ in $\tau$ and quarkonium decays. 
These low-$Q$ measurements are not affected
by gluinos heavier than 10 GeV. Gluino contributions 
to the high-$Q$ data point are of
the same order as the experimental uncertainties. 

With the $\alpha_s$ data and latest hadronic data included, 
the 2010 SUSY PDF fits reduce the allowed range of gluino masses, 
as compared to the 2004 fits \cite{Berger:2004mj}.
For example, Fig.~\ref{res} illustrates SUSY fits with a
constant QCD coupling strength, $\alpha_s(M_Z)=0.118$, for 
gluino masses $m_{\tilde g}$ shown in the figure. 
The left subfigure compares the gluon PDF obtained in SUSY fits with
$m_{\tilde g}=20$ and 50 GeV (solid lines) to the CT10 error band
(corresponding to $m_{\tilde g} =\infty$). It is clear that too light
gluinos distort the shape of the CT10 gluon PDF to an unacceptable
level. The right subfigure shows the differences
$\Delta\chi^2=\chi^2(\alpha_s,m_{\tilde{g}})-\chi^2_{CT10}$.
One can see that $\Delta \chi^2 > 100$ for $m_{\tilde g} <
25$ GeV, suggesting that gluinos lighter than 25 GeV are excluded at
about 90\% C.L., for $\alpha_s(M_Z)=0.118$. This improves the 2004
constraint \cite{Berger:2004mj}, $m_{\tilde g} > 12 $ GeV for
$\alpha_s(M_Z)=0.118$, by a factor of two. Similarly, the 2004
limit on $m_{\tilde g}$ for a free $\alpha_s$, which allowed
$m_{\tilde g} = 1\mbox{ GeV}$ for $\alpha_s(M_Z)=0.135$, is increased to
$m_{\tilde g} > 13$ GeV for any $\alpha_s$ in the 2010 fit. 

Gluinos with
mass about 50 GeV remain allowed both by the global fits and LEP data
analysis. Such light gluinos may alter cross sections for jet
production and other LHC processes \cite{Berger:2010rj}. 
We provide tables of PDFs with contributions of
gluinos in this mass range to explore phenomenological implications.

\shadedfigure[h]{
\caption{\small{Rapidity distributions of $W^+$ and Higgs bosons computed
with LO-MC and NLO PDF's.}} 
\label{Even_gen}
\begin{center}
\includegraphics[width=0.49\textwidth]{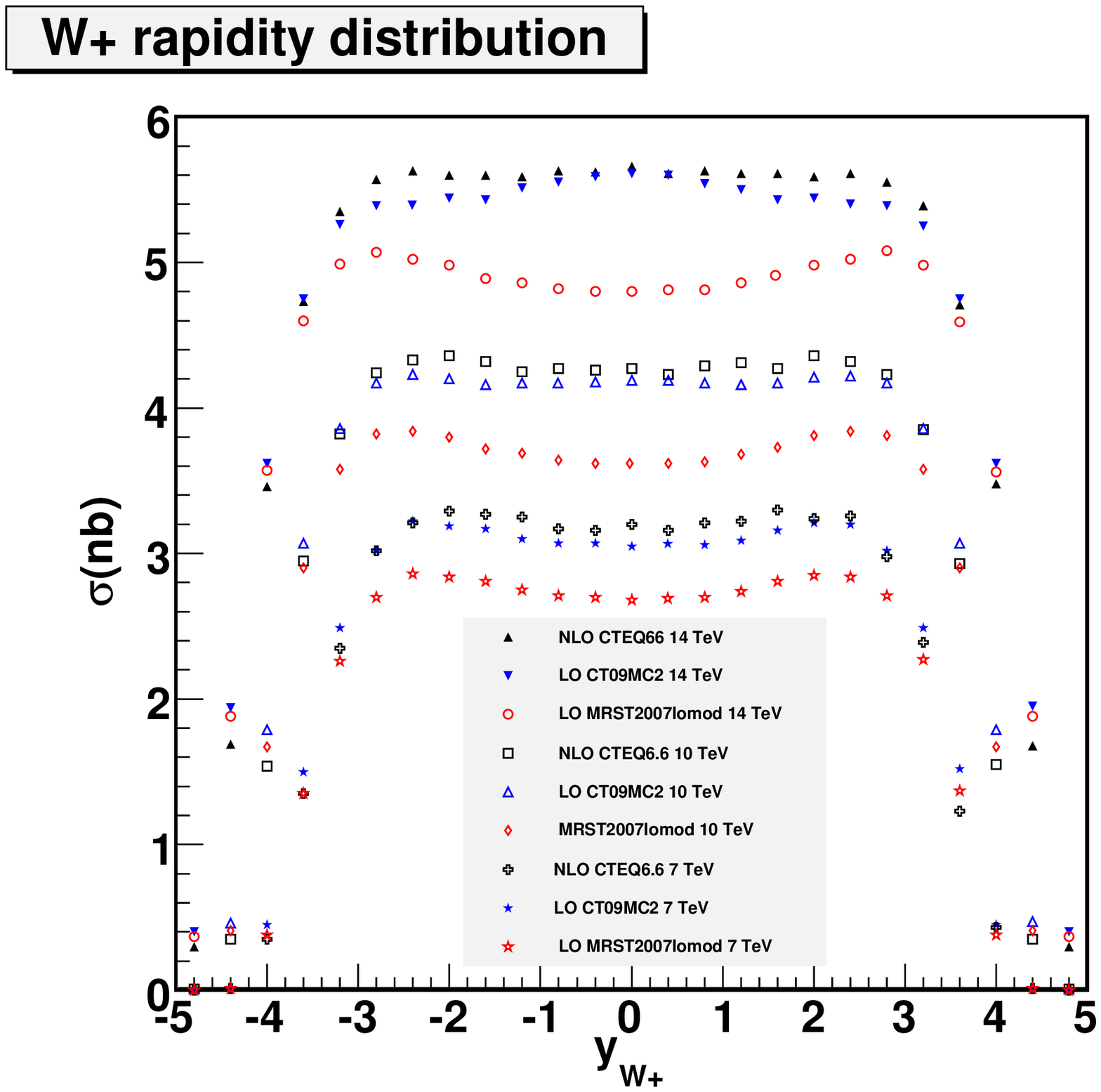}
\includegraphics[width=0.49\textwidth]{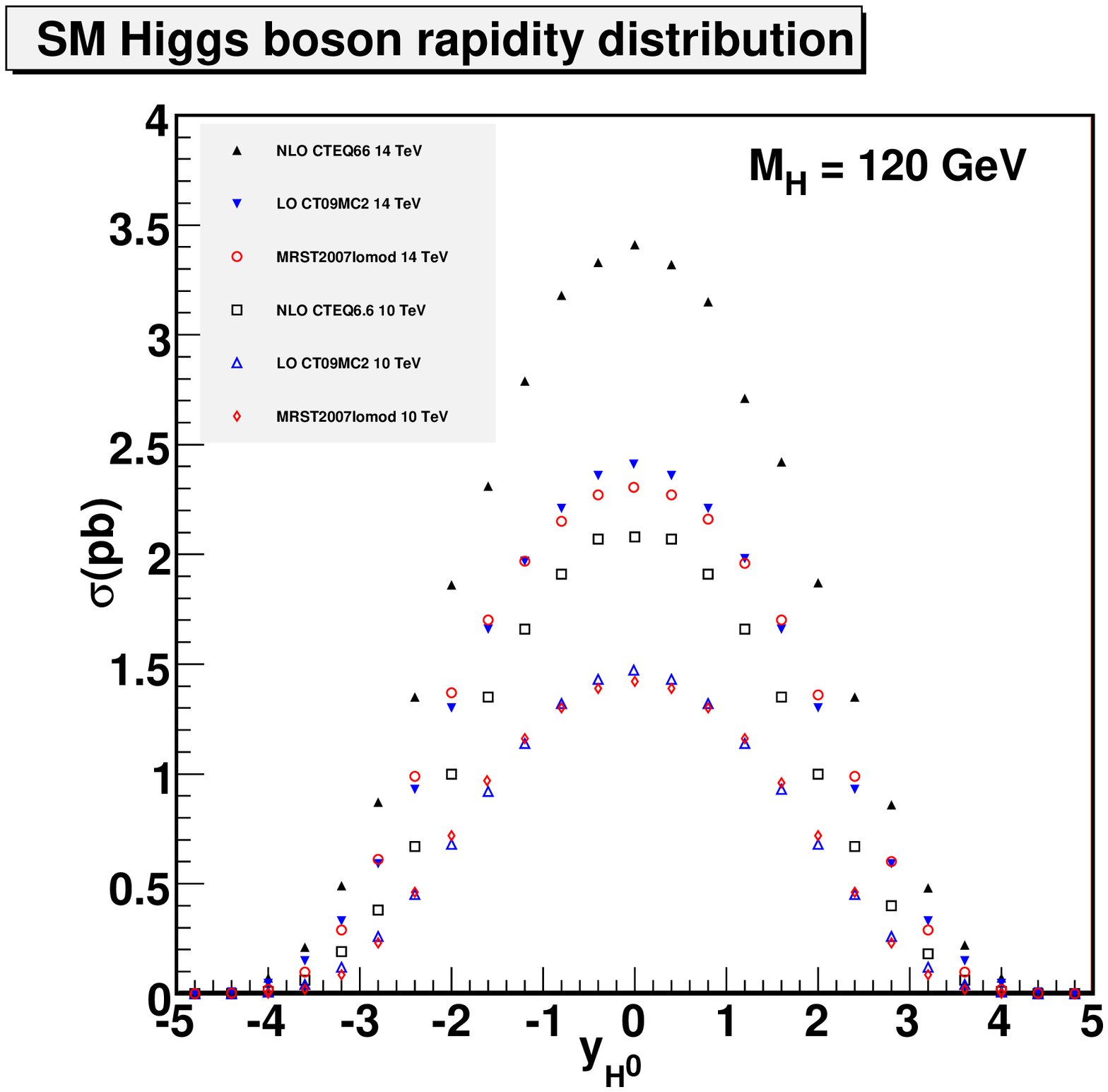}\\
\includegraphics[width=0.49\textwidth]{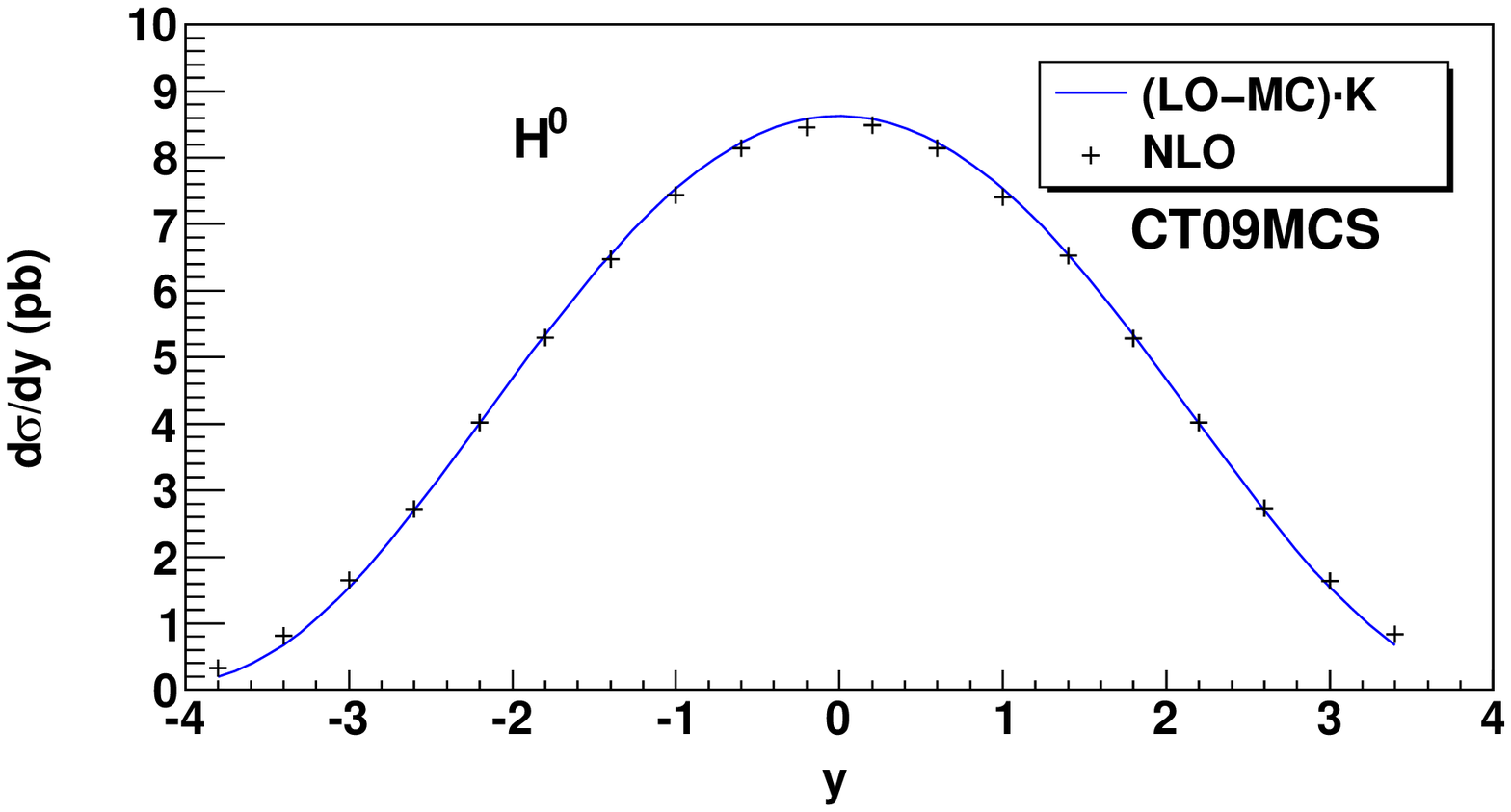}
\includegraphics[width=0.49\textwidth]{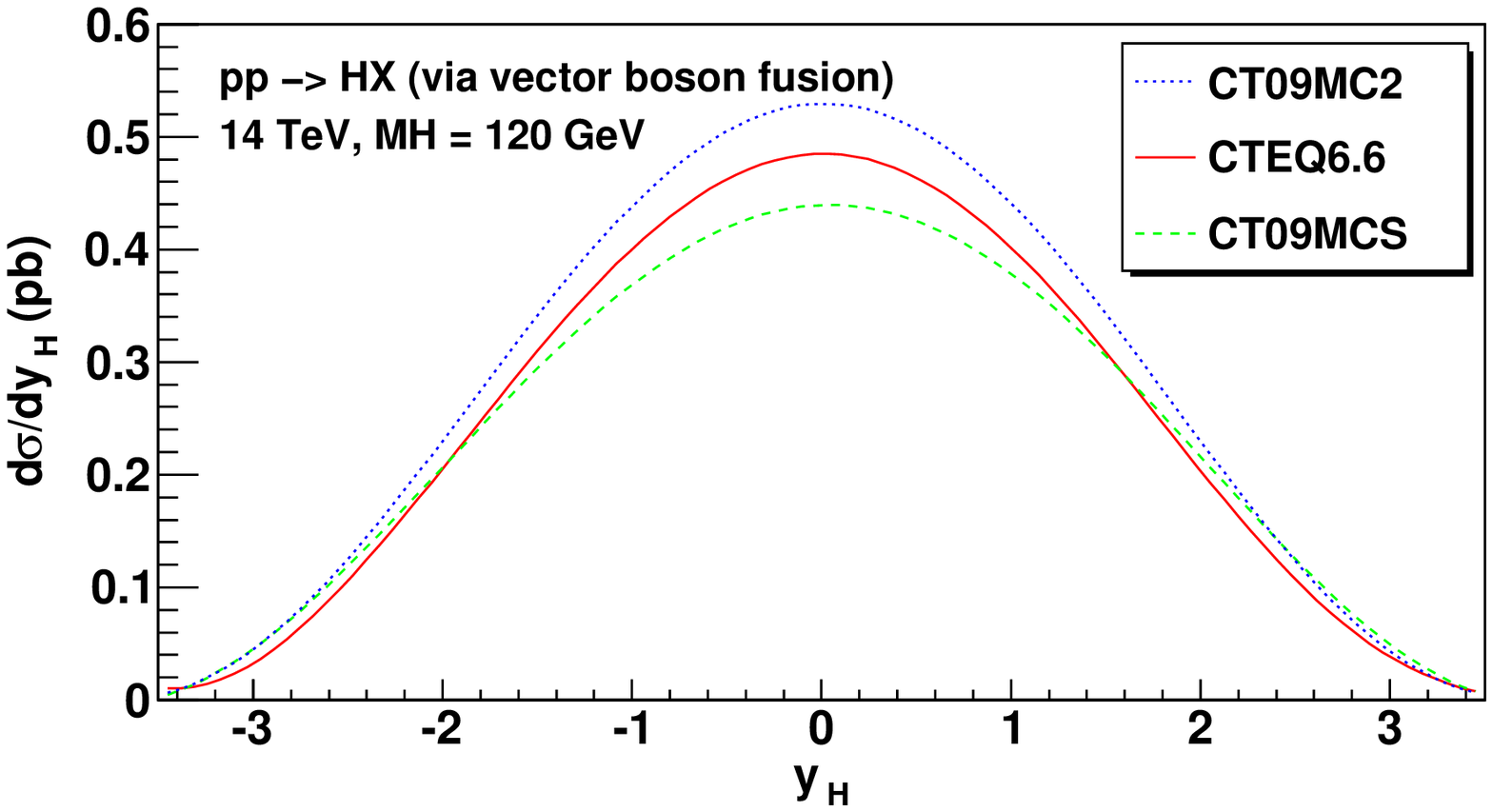}\\
\end{center}
}

\subsection{PDFs for leading-order showering programs}
Monte Carlo event generators, especially,  the most mature 
leading-order generators, play a critical role in all stages 
of modern particle physics. 
Neither conventional LO PDFs, nor NLO PDFs produce satisfactory results
when implemented in the showering programs. In Ref.~\cite{Lai:2009ne},
we modified the usual leading-order global analysis to find optimized PDFs
for leading-order Monte-Carlo (LO-MC) simulations at the LHC.
Besides the usual constraints from the existing 
hard-scattering experimental data, the joint input of this 
analysis incorporates 
pseudodata points for cross sections of
$W$, $Z$, $t\bar t$, SM Higgs, and $gg\rightarrow bb^\prime$
production at the LHC, as {\it predicted} by NLO QCD theory. The PDFs
resulting from this analysis are not strictly at the leading order:
they include some information about key LHC processes evaluated at NLO. 
Event generators, including ``LO event generators",
have some elements of higher-order contributions and, in this sense, 
are not at the stated order in the QCD coupling. We can use their
flexibility to find the LO-MC PDFs that better reproduce 
the benchmark NLO cross sections, when combined with LO matrix
elements in a fixed-order calculation or a LO event generator.

To examine the available possibilities, we provide 
three representative LO-MC PDF sets, designated as CT09MCS, 
CT09MC1, and CT09MC2. These PDFs realize different strategies for
bringing the LO predictions closer to NLO. Their differences
stem from varying assumptions about the running of $\alpha_s$
(evaluated at one or two loops), factorization scales
in the LHC pseudodata cross sections (fixed or fitted), and the
momentum sum rule imposed on the PDFs (exact or relaxed by 10-15\%
\cite{Sherstnev:2007nd}). 
 An example of the LO-MC PDFs in action is shown 
in Fig.\ref{Even_gen}, which compares 
cross sections for $W^+$ boson and Standard Model Higgs boson rapidity
distributions at the LHC, obtained with LO matrix elements and
CT09MC2 and MRST2007lomod PDFs \cite{Sherstnev:2007nd}, and at NLO 
with CTEQ6.6. In the $W^+$ production
case (upper left subfigure), the CT09MC2 calculation closely reproduces both the
normalization and shape of the NLO cross section at all three LHC
energies, while the  MRST2007lomod prediction differs from NLO 
in normalization at $\sqrt{s}=7 $ TeV, and both in normalization and
shape at 10 and 14 TeV. For Higgs production (upper right subfigure), both
CT09MC2 and MRST2007lomod predictions provide almost identical
distributions, which are smaller than the NLO prediction by a nearly
constant normalization factor. This difference with NLO reflects especially
large virtual corrections present in Higgs production cross sections, which
cannot be completely compensated by an increase in
the LO gluon density. However, since the average normalization factors
$K$ for each pseudodata process are also known from the fit (and published in our paper), end users can multiply 
the LO-MC cross sections for Higgs production and other
pseudodata processes by these K-factors to better
approximate the NLO cross sections (cf. the lower left subfigure). If an LHC
process is not included as the pseudodata, comparison of LO
predictions based on several LO-MC sets may still provide a
reasonable estimate of the NLO cross section, as illustrated by 
the cross section for SM Higgs boson production 
via vector boson fusion in the lower
right subfigure.

In summary, the CT10 and CT10W sets are based on the most 
up-to-date information about the PDFs available 
from global hadronic experiments.   
There are 26 free parameters in both new PDF sets; thus, 
there are 26 eigenvector directions and a total of 52 error 
PDFs for both CT10 and CT10W. The CT10 and CT10W  PDF error 
sets, along with the accompanying $\alpha_s$ error sets, 
allow for a complete calculation of the combined 
PDF+$\alpha_s$ uncertainties for any observable.

\subsection*{Acknowledgments} 
This work was supported in part 
by by the U.S. Department of Energy under Grants DE-AC02-06CH11357, 
DE-FG02-04ER41299, and DE-SC0003870;
by the U.S. National Science Foundation under grant PHY-0855561;
by the National Science Council of Taiwan under grants
NSC-98-2112-M-133-002-MY3 and NSC-99-2918-I-133-001;
and by Lightner-Sams Foundation.


\subsection[References]{}

\end{document}